\documentclass[aps,prd,floats,epsf,superscriptaddress,nofootinbib,amsmath]{revtex4}
\usepackage{amsmath,amssymb}
\usepackage{graphicx}
\usepackage{longtable}
\usepackage{multirow}
\newcommand{\specialcellc}[2][c]{%
  \begin{tabular}[#1]{@{}l@{}}#2\end{tabular}}
\newcommand{\specialcellt}[2][t]{%
  \begin{tabular}[#1]{@{}l@{}}#2\end{tabular}}
\allowdisplaybreaks
\LTcapwidth=\textwidth

\begin{document}

{\small
\begin{flushright}
IPMU15-0172
\end{flushright} }
\title{Effective Field Theory of Integrating out Sfermions in the MSSM:\\
Complete One-Loop Analysis}

\author{Ran Huo}
\email[e-mail: ]{ran.huo@ipmu.jp}
\affiliation{Kavli IPMU (WPI), UTIAS, The University of Tokyo, Kashiwa, Chiba 277-8583, Japan}

\begin{abstract}
We apply the covariant derivative expansion of the Coleman-Weinberg potential to the sfermion sector in the minimal supersymmetric standard model, matching it to the relevant dimension-6 operators in the standard model effective field theory at one-loop level. Emphasis is paid to nondegenerate large soft supersymmetry breaking mass squares, and the most general analytical Wilson coefficients are obtained for all pure bosonic dimension-6 operators. In addition to the nonlogarithmic contributions, they generally have another logarithmic contributions. Various numerical results are shown, in particular the constraints in the large $X_t$ branch reproducing the $125$~GeV Higgs mass can be pushed to high values to almost completely probe the low stop mass region at the future FCC-ee experiment, even given the Higgs mass calculation uncertainty.
\end{abstract}

\maketitle

\section{Introduction}

The first task of next generation colliders such as the ILC, the CEPC and the FCC-ee is precision measurement, of Higgs physics as well as Z-pole physics and trigauge boson physics and so on. If there is new physics, it is expected to show up first as corrections to the standard model (SM) processes, subject to such precision measurements. The corrections can be naturally sorted by dimension-6 operators in the SM effective field theory (EFT), which has a total of 59 independent operators for one family of fermions as a complete basis~\cite{Grzadkowski:2010es}. Precision measurements can be translated into a set of constraints on (part of) the operators.

The operator level fitting process is (UV) model independent. On the other hand, in literature there are many popular new physics models with various motivations and merits. Previously the matching of new physics to the model independent operator constraints are mostly electroweak precision test (EWPT) and Higgs cross sections/decay branching ratios. For the most important subset of pure bosonic operators and at one-loop level which is usually leading with occasional exceptions, a method called the covariant derivative expansion (CDE)~\cite{Henning:2014gca,Henning:2014wua} has greatly facilitate the matching procedure, with advantage of being complete for matching all operators simultaneously, model independence in computation process, and giving analytical results.

One can match any desirable UV models to the SM EFT~\cite{Henning:2014gca,Henning:2014wua,Drozd:2015kva,Chiang:2015ura,Huo:2015exa}, but before that some generalizations of the CDE technique need to be made. In addition to the generalization to fermionic degree of freedom (DOF) which has extra gamma matrices related contributions~\cite{Henning:2014wua,Huo:2015exa}, for realistic model parameters the original treatment of only degenerated large scales in~\cite{Henning:2014gca,Henning:2014wua} is based on an oversimplified assumption to be dropped. Since the CDE has already been formulated in matrix basis, it is not a difficult generalization to use nondegenerate large scales, and use heavily the integration of Feynman parametrization at textbook level after taking traces.

Here we do sample calculations for the sfermion sector of the minimal supersymmetric standard model (MSSM). We expand the Coleman-Weinberg (CW) potential with covariant derivatives, matching it to a maximal set of bosonic operators, getting their Wilson coefficients. In addition to the nondegeneracy of soft mass squares we keep the small bottom/tau Yukawa couplings, and the differences between the squark sector and the slepton sector are accounted by different $SU(3)_c$ representations and $U(1)_Y$ hypercharges. The analytical results at one-loop level should be complete and realistic to use for various purposes.

A lot of works have been done in the indirect precision constraints of the squark sector of the MSSM, including the one-loop level Feynman diagram calculation. To provide some new useful numerics here we consider the MSSM Higgs mass constraint in detail, in particular we solve the mixing term $X_t$ for the required SM like Higgs mass. There are in general two solutions for a required SM like Higgs mass, and we find the precision constraints are especially effective for the large $X_t$ branch.

This paper is organized as follows. In Sec.~\ref{sec:for}, we briefly review the dimension-6 operator basis and the bosonic CDE formulism, then we discuss the structure needed for large mass nondegeneracy. Next in Sec.~\ref{sec:sferm} we show the sfermion sector input to the CDE formulism, and then the main analytical results. Sec.~\ref{sec:num} is devoted to various numerical applications of the analytical results. We conclude in Sec.~\ref{sec:dis}. At last, Appendix~\ref{sec:app} provides a list of useful auxiliary formulas for the integration of Feynman parameters.

\section{Formulism\label{sec:for}}

\begin{table}[th]
\begin{tabular}{cccccccc}
\hline\hline
~Symbol~ & Operator expression &~~& ~Symbol~ & Operator expression &~~& ~Symbol~ & Operator expression\\
\hline
${\cal O}_6$ & $(H^\dag H)^3$ && ${\cal O}_{GG}$ & $g_s^2H^\dag HG^a_{\mu\nu}G^{a\mu\nu}$ && ${\cal O}_{W}$ & $ig(H^\dag\overleftrightarrow{D}_\mu t^a H )D_\nu W^{a\mu\nu}$ \\
${\cal O}_H$ & $\frac12(\partial_\mu (H^\dag H))^2$ && ${\cal O}_{WW}$ & $g^2H^\dag HW^a_{\mu\nu}W^{a\mu\nu}$ && ${\cal O}_{B}$ & $ig'(H^\dag\overleftrightarrow{D}_\mu H)\partial_\nu B^{\mu\nu}$ \\
${\cal O}_T$ & $\frac12(H^\dag \overleftrightarrow{D}_\mu H)^2$ && ${\cal O}_{BB}$ & $g'^2H^\dag HB_{\mu\nu}B^{\mu\nu}$ && ${\cal O}_{HW}$ & $2ig(D_\mu H)^\dag t^a(D_\nu H)W^{a\mu\nu}$ \\
${\cal O}_R$ & $(H^\dag H)(D_\mu H^\dag D^\mu H)$ && ${\cal O}_{WB}$ & $2gg'H^\dag t^a HW_{\mu\nu}^a B^{\mu\nu}$ && ${\cal O}_{HB}$ & $2ig'Y_H(D_\mu H)^\dag(D_\nu H)B^{\mu\nu}$ \\
 && & && & ${\cal O}_D$ & $(D_\mu D^\mu H^\dag)(D_\nu D^\nu H)$\\
\hline\hline
\end{tabular}
\caption{Independent CP-even dimension-6 operators composed of only the Higgs and gauge boson fields that are relevant to the analysis in this work.
\label{tab:Op}}
\end{table}

The operator basis we use is listed in Table~\ref{tab:Op}, which includes only Higgs and electroweak gauge boson fields. The basis is natural in sense that it directly captures all the possibilities generated by the CDE formulism, before removing any redundant operators.

The bosonic CDE formulism given in~\cite{Huo:2015exa} can be summarized as the dimension-6 terms generated by the one-loop integration
\begin{equation}
\mathcal{L}_{\text{CDE}}=\frac{n_B}{2}\int_0^\infty du\int\frac{d^dp_E}{(2\pi)^d}\sum_{m=1}^\infty(-1)^m\text{tr}\bigg[\bigg(\frac{1}{p_E^2+M^2+u}\Big[\delta\tilde{V}''+\tilde{G}\Big]\bigg)^m\frac{1}{p_E^2+M^2+u}\bigg]~,\label{eq:CDE}
\end{equation}
where
\begin{align}
\delta\tilde{V}''=&e^{-iD\frac{\partial}{\partial p}}\delta V''e^{iD\frac{\partial}{\partial p}}=\delta V''+\sum_{n=1}^\infty\frac{(-i)^n}{n!}\thinspace D_{\mu_1}\cdots D_{\mu_n}\delta V''\enspace{\textstyle\frac{\partial}{\partial p}^{\mu_1}\cdots\frac{\partial}{\partial p}^{\mu_n}},\\
\tilde{G}=&gp^\mu t^a\bigg(iF_{\nu\mu}^a{\textstyle\frac{\partial}{\partial p}^\nu}+\frac{4}{3!}D_\rho F_{\nu\mu}^a{\textstyle\frac{\partial}{\partial p}^\rho\frac{\partial}{\partial p}^\nu}+\cdots\bigg)+gt^a\bigg(\frac{2}{3!}D^\mu F_{\nu\mu}^a{\textstyle\frac{\partial}{\partial p}^\nu}+\cdots\bigg)+g^2t^at^b\bigg(\frac{1}{4}F^a_{\nu\mu}F^{b\rho\mu}{\textstyle\frac{\partial}{\partial p}^\nu\frac{\partial}{\partial p}_\rho}+\cdots\bigg).\label{eq:V&G}
\end{align}
In Eq.~(\ref{eq:CDE}) $n_B$ is the DOF for each entry, being $1$ for a real bosonic DOF and $2$ for a complex bosonic DOF. The dimension-2 auxiliary number $u$ is introduced as a trick to regularize orders of commutators of $\frac{\partial}{\partial p}$s acting on $(p^2-M^2)$s~\cite{Zuk:1985sw}. A Wick rotation is performed and subscript ``E'' indicates Euclidean. The CDE assumes in the CW potential the double derivatives (indicated by $'$) to the beyond SM fields acting on the Lagrangian potential terms can be decomposed as $V''=M^2+\delta V''$, where $M^2$ is some large constant squared mass term which is irrelevant to Higgs vacuum expectation value (VEV), and $\delta V''$ on the other hand captures the spacetime dependent part (namely the SM Higgs) of potential terms, with not only the physical Higgs boson but the Nambu-Goldstone modes before electroweak symmetry breaking. At last $\tilde{\enspace}$ generally indicates a Baker-Campbell-Hausdorff expansion with covariant derivatives and $\frac{\partial}{\partial p}$s, which can be understand as introducing spacetime dependence in the momentum space~\cite{Chan:1985ny} (and the covariant generalization in gauge field $\partial\to D=\partial-igA$ is given in~\cite{Gaillard:1985uh,Cheyette:1987qz}, in which the $\tilde{G}$ terms arise).

Note that the $\delta\tilde{V}''$ and $\tilde{G}$ are all matrices, with each entry coupling to a gauge DOF of new particle to be integrated out. So is the large spacetime independent mass term $M^2$, which in all known models is a diagonal matrix (if $M^2$ is not diagonal the new physics has intrinsic mixing irrelevant to the SM, such a model seems difficult to get motivated). This directly shows the way of generalizing the CDE formulism with nondegenerate large mass parameters, namely treating $1/(p_E^2+M^2+u)$ as a diagonal matrix $(p_E^2+M^2+u)^{-1}$ and doing the multiplication with full respect to their matrix nature and noncommutativity with the $(\delta\tilde{V}''+\tilde{G})$s. The $\textstyle\frac{\partial}{\partial p}$ action on any $(p_E^2+M^2+u)^{-1}$ factor (as well as the factor $p$ in the first term of $\tilde{G}$) will give a commutative $p$ number on the numerator and introduce no ambiguity. Eventually the overall trace operation reduces the matrix to numbers.

The later jobs such as the textbook trick of Feynman parametrization for each term, the dimensional regularization/reduction integration over $\int d^dp_E$ and the integration over $\int du$ (with the boundary $u\to\infty$ dropped as $\overline{\text{MS}}$ subtraction) are quite tedious, but possible especially with a symbolic calculation tool such as the \texttt{Mathematica}.

\section{The SFermion Sector\label{sec:sferm}}

In the following we do the CDE matching for the squark sector explicitly. Without specifying the $Y_q,Y_t,Y_b$ values the slepton sector results can be correspondingly obtained, though one should keep in mind that the $c_{GG}$ vanishes and every other Wilson coefficient should be divided by a factor of $3$ as the $SU(3)_c$ color multiplicity.

The CW potential matrix in the basis of $(\tilde{t}_L,\tilde{b}_L,\tilde{t}_R,\tilde{b}_R)$ has large spacetime independent masses (suppressing the $SU(3)_c$ components)
\begin{equation}
M^2=\left(\begin{array}{cccc}
M_{\tilde{q}}^2 & 0 & 0 & 0 \\
0 & M_{\tilde{q}}^2 & 0 & 0 \\
0 & 0 & M_{\tilde{t}}^2 & 0 \\
0 & 0 & 0 & M_{\tilde{b}}^2
\end{array}\right),\label{eq:M^2}
\end{equation}
and the spacetime dependent terms can be written in a decomposition $\delta V''=\delta V''_F+\delta V''_D+\delta V''_X$, with
\begin{align}
\delta V''_F&=\left(\begin{array}{cccc}
y_t^2H^{0\ast}H^0+y_b^2H^-H^+ & (-y_t^2+y_b^2)H^{0\ast}H^+ & 0 & 0 \\
(-y_t^2+y_b^2)H^-H^0 & y_t^2H^{0\ast}H^0+y_b^2H^-H^+ & 0 & 0 \\
0 & 0 & y_t^2(H^{0\ast}H^0+H^-H^+) & 0 \\
0 & 0 & 0 & y_b^2(H^{0\ast}H^0+H^-H^+)
\end{array}\right),\label{eq:VF}\\
\delta V''_D&=\cos2\beta\left(\begin{array}{cccc}
(\frac{g^2}{4}\hspace{-0.3em}-\hspace{-0.3em}\frac{Y_qg'^2}{2})H^{0\ast}H^0\hspace{-0.3em}-\hspace{-0.3em}(\frac{g^2}{4}\hspace{-0.3em}+\hspace{-0.3em}\frac{Y_qg'^2}{2})H^-H^+ & \hspace{-4em}-\frac{g^2}{2}H^{0\ast}H^+ & \hspace{-4em}0 & \hspace{-4em}0 \\
\hspace{-5em}-\frac{g^2}{2}H^-H^0 & \hspace{-5em}-(\frac{g^2}{4}\hspace{-0.3em}+\hspace{-0.3em}\frac{Y_qg'^2}{2})H^{0\ast}H^0\hspace{-0.3em}+\hspace{-0.3em}(\frac{g^2}{4}\hspace{-0.3em}-\hspace{-0.3em}\frac{Y_qg'^2}{2})H^-H^+ & \hspace{-4em}0 & \hspace{-4em}0 \\
\hspace{-5em}0 & \hspace{-4em}0 & \hspace{-5em}\frac{Y_tg'^2}{2}(H^{0\ast}H^0+H^-H^+) & \hspace{-4em}0 \\
\hspace{-5em}0 & \hspace{-4em}0 & \hspace{-4em}0 & \hspace{-3em}\frac{Y_bg'^2}{2}(H^{0\ast}H^0+H^-H^+)
\end{array}\right),\label{eq:VD}\\
\delta V''_X&=\left(\begin{array}{cccc}
0 & 0 & y_tH^{0\ast}X_t & y_bH^+X_b \\
0 & 0 & -y_tH^-X_t & y_bH^0X_b \\
y_tH^0X_t & -y_tH^+X_t & 0 & 0 \\
y_bH^-X_b & y_bH^{0\ast}X_b & 0 & 0
\end{array}\right),\label{eq:VX}
\end{align}
which count the supersymmetric F-term, D-term and the trilinear $X$-term contributions respectively. Here the Yukawa couplings are defined as their SM values such as $y_t=\sqrt{2}m_t/v=y_t^\text{MSSM}\sin\beta$ with $v=246$~GeV, and we have ignored any possible \emph{CP} phases. Plugging them into Eq.~(\ref{eq:CDE}) and collecting the dimension-6 operators as described in the appendix of~\cite{Huo:2015exa}, we tabulate the resulting Wilson coefficient for each operator in the following long table.

\begin{longtable}{|c|p{0.49\linewidth}|p{0.4\linewidth}|}
\caption{The Wilson coefficients of integrating out the sfermion sector. The operators are listed in Table~\ref{tab:Op} and the Wilson coefficients are defined with dimension-$-2$.}
\\
\hline
 & Nonlogarithmic Contributions & Logarithmic Contributions \\
\endfirsthead
\caption{continued}
\\
\hline
 & Nonlogarithmic Contributions & Logarithmic Contributions \\
\endhead
\hline
$(4\pi)^2c_6$ &
\specialcellc{$-\frac{1}{32M_{\tilde{q}}^2}$
\specialcellt{$\Big((y_t^2+y_b^2-g'^2 Y_q\cos2\beta)\big(16(y_t^4+y_b^4-y_b^2 y_t^2)$ \\
\hspace{-1em}$+(12g^2(y_t^2-y_b^2)-8g'^2Y_q(y_b^2+y_t^2))\cos2\beta$ \\
\hspace{-1em}$+(3g^4+4g'^4Y_q^2)\cos^22\beta\big)\Big)$} \\
$-\frac{1}{16M_{\tilde{t}}^2}\Big(2y_t^2+g'^2Y_t\cos2\beta\Big)^3-(t\leftrightarrow b)$ \\
$+\frac{3y_t^2X_t^2}{32M_{\tilde{q}}^2 M_{\tilde{t}}^2 (M_{\tilde{q}}^2-M_{\tilde{t}}^2)^2}$\specialcellt{$\Big(16(M_{\tilde{q}}^2-M_{\tilde{t}}^2)^2y_t^4$ \\
\hspace{-6em}$-8(M_{\tilde{q}}^2-M_{\tilde{t}}^2)y_t^2(g^2M_{\tilde{t}}^2-2g'^2(M_{\tilde{q}}^2 Y_t+M_{\tilde{t}}^2 Y_q))\cos2\beta$ \\
\hspace{-6em}$+\big(g^4M_{\tilde{t}}^2(M_{\tilde{q}}^2+M_{\tilde{t}}^2)-4g^2g'^2M_{\tilde{t}}^2(M_{\tilde{q}}^2(Y_q+2Y_t)+M_{\tilde{t}}^2Y_q)$ \\
\hspace{-6em}$+4g'^4(M_{\tilde{q}}^4Y_t^2+M_{\tilde{q}}^2M_{\tilde{t}}^2(Y_q^2+4Y_qY_t+Y_t^2)+M_{\tilde{t}}^4Y_q^2)\big)$\\
\hspace{-5em}$\times\cos^22\beta\Big)+(t\leftrightarrow b,g^2\to-g^2)$} \\
$-\frac{3y_t^4X_t^4}{8M_{\tilde{q}}^2M_{\tilde{t}}^2(M_{\tilde{q}}^2-M_{\tilde{t}}^2)^3}$\specialcellt{$\Big(4(M_{\tilde{q}}^4-M_{\tilde{t}}^4)y_t^2$\\
\hspace{-5.5em}$-\big(g^2(5M_{\tilde{q}}^2M_{\tilde{t}}^2+M_{\tilde{t}}^4)-2g'^2(M_{\tilde{q}}^4Y_t+M_{\tilde{t}}^4Y_q$ \\
\hspace{-5.5em}$+5 M_{\tilde{q}}^2 M_{\tilde{t}}^2 (Y_q+Y_t))\big)\cos2\beta\Big)-(t\leftrightarrow b,g^2\to-g^2)$} \\
$+\frac{y_t^6X_t^6}{2M_{\tilde{q}}^2M_{\tilde{t}}^2(M_{\tilde{q}}^2-M_{\tilde{t}}^2)^4}\Big(M_{\tilde{q}}^4+10 M_{\tilde{q}}^2 M_{\tilde{t}}^2+M_{\tilde{t}}^4\Big)+(t\leftrightarrow b)$
} &
\specialcellc{$+\frac{3y_t^2X_t^2}{16(M_{\tilde{q}}^2-M_{\tilde{t}}^2)^3}$\specialcellt{$\Big(\big(g^2-2g'^2(Y_q+Y_t)\big)\big(4(M_{\tilde{q}}^2-M_{\tilde{t}}^2)y_t^2$ \\
\hspace{-3.5em}$-(g^2M_{\tilde{t}}^2-2g'^2(M_{\tilde{q}}^2Y_t+M_{\tilde{t}}^2 Y_q))\cos2\beta$\big)\\
\hspace{-3em}$\times\cos2\beta\Big)\ln\frac{M_{\tilde{q}}^2}{M_{\tilde{t}}^2}+(t\leftrightarrow b,g^2\to-g^2)$} \\
$+\frac{3y_t^4X_t^4}{4(M_{\tilde{q}}^2-M_{\tilde{t}}^2)^4}$\specialcellt{$\Big(4(M_{\tilde{q}}^2-M_{\tilde{t}}^2)y_t^2-\big(g^2(M_{\tilde{q}}^2+2M_{\tilde{t}}^2)$ \\
\hspace{-3em}$-2g'^2(M_{\tilde{q}}^2(Y_q+2Y_t)+M_{\tilde{t}}^2(2Y_q+Y_t))\big)$ \\
\hspace{-3em}$\times\cos2\beta\Big)\ln\frac{M_{\tilde{q}}^2}{M_{\tilde{t}}^2}+(t\leftrightarrow b,g^2\to-g^2)$} \\
$-\frac{3y_t^6X_t^6}{(M_{\tilde{q}}^2-M_{\tilde{t}}^2)^5}\Big(M_{\tilde{q}}^2+M_{\tilde{t}}^2\Big)\ln\frac{M_{\tilde{q}}^2}{M_{\tilde{t}}^2}-(t\leftrightarrow b)$
} \\
\hline\hline
$(4\pi)^2c_H$ &
\specialcellc{$+\frac{1}{4M_{\tilde{q}}^2}\Big(y_t^2+y_b^2-g'^2Y_q\cos2\beta\Big)^2$ \\
$+\frac{1}{8M_{\tilde{t}}^2}\Big(2y_t^2+g'^2Y_t\cos2\beta\Big)^2+(t\leftrightarrow b)$ \\
$-\frac{y_t^2X_t^2}{8M_{\tilde{q}}^2M_{\tilde{t}}^2(M_{\tilde{q}}^2-M_{\tilde{t}}^2)^3}$\specialcellt{$\Big(8M_{\tilde{q}}^6 y_t^2-2M_{\tilde{q}}^4M_{\tilde{t}}^2(13y_t^2-y_b^2)$ \\
\hspace{-5.5em}$-4M_{\tilde{t}}^6(y_t^2+y_b^2)+2M_{\tilde{q}}^2 M_{\tilde{t}}^4(17y_t^2-5y_b^2)$ \\
\hspace{-5.5em}$+\big(3g^2M_{\tilde{q}}^2M_{\tilde{t}}^2(M_{\tilde{q}}^2+5 M_{\tilde{t}}^2)+4g'^2(M_{\tilde{t}}^6Y_q+M_{\tilde{q}}^6 Y_t$ \\
\hspace{-5.5em}$-M_{\tilde{q}}^2M_{\tilde{t}}^4(5Y_q+2Y_t)-M_{\tilde{q}}^4M_{\tilde{t}}^2(2Y_q+5Y_t))\big)\cos2\beta\Big)$ \\
\hspace{-5.5em}$-(t\leftrightarrow b,g^2\to-g^2)$} \\
$+\frac{y_t^4X_t^4}{4M_{\tilde{q}}^2M_{\tilde{t}}^2(M_{\tilde{q}}^2-M_{\tilde{t}}^2)^4}$\specialcellt{$\Big(2 M_{\tilde{q}}^6-15 M_{\tilde{q}}^4 M_{\tilde{t}}^2-24 M_{\tilde{q}}^2M_{\tilde{t}}^4+M_{\tilde{t}}^6\Big)$ \\
\hspace{-5.5em}$+(t\leftrightarrow b)$} \\
$+\frac{y_t^2y_b^2X_t^2X_b^2}{4M_{\tilde{q}}^2(M_{\tilde{q}}^2-M_{\tilde{t}}^2)^3(M_{\tilde{q}}^2-M_{\tilde{b}}^2)^3}$\specialcellt{$\Big(2M_{\tilde{q}}^8-M_{\tilde{q}}^4(M_{\tilde{t}}^4+M_{\tilde{b}}^4)+2M_{\tilde{t}}^4M_{\tilde{b}}^4$ \\
\hspace{-8.2em}$+5(M_{\tilde{q}}^4+M_{\tilde{t}}^2M_{\tilde{b}}^2)M_{\tilde{q}}^2(M_{\tilde{t}}^2+M_{\tilde{b}}^2)-22M_{\tilde{q}}^4M_{\tilde{t}}^2M_{\tilde{b}}^2\Big)$}
}
&
\specialcellc{$+\frac{3y_t^2X_t^2M_{\tilde{t}}^2}{4(M_{\tilde{q}}^2-M_{\tilde{t}}^2)^4}$\specialcellt{$\Big(2 M_{\tilde{t}}^2(y_t^2-y_b^2)+\big(g^2 (2 M_{\tilde{q}}^2+M_{\tilde{t}}^2)$ \\
\hspace{-3em}$-4g'^2M_{\tilde{q}}^2(Y_q+Y_t)\big)\cos2\beta\Big)\ln\frac{M_{\tilde{q}}^2}{M_{\tilde{t}}^2}$ \\
\hspace{-3em}$+(t\leftrightarrow b,g^2\to-g^2)$} \\
$+\frac{3y_t^4X_t^4M_{\tilde{t}}^2}{2(M_{\tilde{q}}^2-M_{\tilde{t}}^2)^5}\Big(5M_{\tilde{q}}^2+M_{\tilde{t}}^2\Big)\ln\frac{M_{\tilde{q}}^2}{M_{\tilde{t}}^2}+(t\leftrightarrow b)$ \\
$-\frac{3y_t^2y_b^2X_t^2X_b^2}{2(M_{\tilde{q}}^2-M_{\tilde{t}}^2)^4(M_{\tilde{q}}^2-M_{\tilde{b}}^2)^4(M_{\tilde{t}}^2-M_{\tilde{b}}^2)}$\specialcellt{$\Big[\Big((M_{\tilde{q}}^8-M_{\tilde{t}}^4M_{\tilde{b}}^4)$ \\
\hspace{-10.8em}$\times(M_{\tilde{t}}^4-M_{\tilde{b}}^4)-4M_{\tilde{q}}^2M_{\tilde{t}}^2M_{\tilde{b}}^2\big(M_{\tilde{t}}^2(M_{\tilde{q}}^4+M_{\tilde{b}}^4)$\\
\hspace{-10.8em}$-M_{\tilde{b}}^2(M_{\tilde{q}}^4+M_{\tilde{t}}^4)\big)\Big)\ln M_{\tilde{q}}^2$ \\
\hspace{-10.8em}$-(M_{\tilde{q}}^2-M_{\tilde{b}}^2)M_{\tilde{t}}^4\ln M_{\tilde{t}}^2$ \\
\hspace{-10.8em}$+(M_{\tilde{q}}^2-M_{\tilde{t}}^2)M_{\tilde{b}}^4\ln M_{\tilde{b}}^2\Big]$}
}
\\
\hline
$(4\pi)^2c_T$ &
\specialcellc{$+\frac{1}{16M_{\tilde{q}}^2}\Big(2y_t^2-2y_b^2+g^2\cos2\beta\Big)^2$ \\
$-\frac{y_t^2X_t^2}{8M_{\tilde{q}}^2(M_{\tilde{q}}^2-M_{\tilde{t}}^2)^3}$ \specialcellt{$\Big((2y_t^2-2y_b^2+g^2\cos2\beta)$ \\
\hspace{-4.5em}$\times(M_{\tilde{q}}^4-5 M_{\tilde{q}}^2M_{\tilde{t}}^2-2M_{\tilde{t}}^4)\Big)-(t\leftrightarrow b,g^2\to-g^2)$} \\
$+\frac{y_t^4X_t^4}{4M_{\tilde{q}}^2(M_{\tilde{q}}^2-M_{\tilde{t}}^2)^4}\Big(M_{\tilde{q}}^4+10M_{\tilde{q}}^2M_{\tilde{t}}^2+M_{\tilde{t}}^4\Big)+(t\leftrightarrow b)$ \\
$-\frac{y_t^2y_b^2X_t^2X_b^2}{4M_{\tilde{q}}^2(M_{\tilde{q}}^2-M_{\tilde{t}}^2)^3(M_{\tilde{q}}^2-M_{\tilde{b}}^2)^3}$\specialcellt{$\Big(2M_{\tilde{q}}^8-M_{\tilde{q}}^4(M_{\tilde{t}}^4+M_{\tilde{b}}^4)+2M_{\tilde{t}}^4M_{\tilde{b}}^4$ \\
\hspace{-8.2em}$+5(M_{\tilde{q}}^4+M_{\tilde{t}}^2M_{\tilde{b}}^2)M_{\tilde{q}}^2(M_{\tilde{t}}^2+M_{\tilde{b}}^2)-22M_{\tilde{q}}^4M_{\tilde{t}}^2M_{\tilde{b}}^2\Big)$}
}
&
\specialcellc{$-\frac{3y_t^2X_t^2M_{\tilde{t}}^4}{4(M_{\tilde{q}}^2-M_{\tilde{t}}^2)^4}$\specialcellt{$\Big(2y_t^2-2y_b^2+g^2\cos2\beta\Big)\ln\frac{M_{\tilde{q}}^2}{M_{\tilde{t}}^2}$ \\
\hspace{-3em}$-(t\leftrightarrow b,g^2\to-g^2)$} \\
$-\frac{3y_t^4X_t^4M_{\tilde{t}}^2}{2(M_{\tilde{q}}^2-M_{\tilde{t}}^2)^5}\Big(M_{\tilde{q}}^2+M_{\tilde{t}}^2\Big)\ln\frac{M_{\tilde{q}}^2}{M_{\tilde{t}}^2}-(t\leftrightarrow b)$ \\
$+\frac{3y_t^2y_b^2X_t^2X_b^2}{2(M_{\tilde{q}}^2-M_{\tilde{t}}^2)^4(M_{\tilde{q}}^2-M_{\tilde{b}}^2)^4(M_{\tilde{t}}^2-M_{\tilde{b}}^2)}$\specialcellt{$\Big[\Big((M_{\tilde{q}}^8-M_{\tilde{t}}^4M_{\tilde{b}}^4)$ \\
\hspace{-10.8em}$\times(M_{\tilde{t}}^4-M_{\tilde{b}}^4)-4M_{\tilde{q}}^2M_{\tilde{t}}^2M_{\tilde{b}}^2\big(M_{\tilde{t}}^2(M_{\tilde{q}}^4+M_{\tilde{b}}^4)$\\
\hspace{-10.8em}$-M_{\tilde{b}}^2(M_{\tilde{q}}^4+M_{\tilde{t}}^4)\big)\Big)\ln M_{\tilde{q}}^2$ \\
\hspace{-10.8em}$-(M_{\tilde{q}}^2-M_{\tilde{b}}^2)M_{\tilde{t}}^4\ln M_{\tilde{t}}^2$ \\
\hspace{-10.8em}$+(M_{\tilde{q}}^2-M_{\tilde{t}}^2)M_{\tilde{b}}^4\ln M_{\tilde{b}}^2\Big]$}
}
\\
\hline
$(4\pi)^2c_R$ &
\specialcellc{$+\frac{1}{8M_{\tilde{q}}^2}\Big(2y_t^2-2y_b^2+g^2\cos2\beta\Big)^2$ \\
$+\frac{y_t^2X_t^2}{8M_{\tilde{q}}^2 (M_{\tilde{q}}^2-M_{\tilde{t}}^2)^3}$\specialcellt{$
\Big(8M_{\tilde{t}}^4(y_t^2-y_b^2)-4M_{\tilde{q}}^2M_{\tilde{t}}^2(7y_t^2+5y_b^2)$ \\
\hspace{-4.5em}$+4M_{\tilde{q}}^4(11y_t^2+y_b^2)-\big(g^2(5 M_{\tilde{q}}^4+5M_{\tilde{q}}^2M_{\tilde{t}}^2-4M_{\tilde{t}}^4)$ \\
\hspace{-4.5em}$-6g'^2M_{\tilde{q}}^2(M_{\tilde{q}}^2(Y_q+5Y_t)+M_{\tilde{t}}^2(5Y_q+Y_t))\big)\cos2\beta\Big)$ \\
\hspace{-4.5em}$+(t\leftrightarrow b,g^2\to-g^2)$} \\
$-\frac{y_t^4X_t^4}{2M_{\tilde{q}}^2(M_{\tilde{q}}^2-M_{\tilde{t}}^2)^4}\Big(17M_{\tilde{q}}^4+8M_{\tilde{q}}^2M_{\tilde{t}}^2-M_{\tilde{t}}^4\Big)-(t\leftrightarrow b)$ \\
$-\frac{y_t^2y_b^2X_t^2X_b^2}{2M_{\tilde{q}}^2(M_{\tilde{q}}^2-M_{\tilde{t}}^2)^3(M_{\tilde{q}}^2-M_{\tilde{b}}^2)^3}$\specialcellt{$\Big(2M_{\tilde{q}}^8-M_{\tilde{q}}^4(M_{\tilde{t}}^4+M_{\tilde{b}}^4)+2M_{\tilde{t}}^4M_{\tilde{b}}^4$ \\
\hspace{-8.2em}$+5(M_{\tilde{q}}^4+M_{\tilde{t}}^2M_{\tilde{b}}^2)M_{\tilde{q}}^2(M_{\tilde{t}}^2+M_{\tilde{b}}^2)-22M_{\tilde{q}}^4M_{\tilde{t}}^2M_{\tilde{b}}^2\Big)$}
}
&
\specialcellc{$-\frac{3y_t^2X_t^2}{4(M_{\tilde{q}}^2-M_{\tilde{t}}^2)^4}$\specialcellt{$\Big(4M_{\tilde{q}}^4y_t^2-4M_{\tilde{t}}^4y_b^2$ \\
\hspace{-3em}$-\big(g^2(2M_{\tilde{q}}^2M_{\tilde{t}}^2-M_{\tilde{t}}^4)-2g'^2(M_{\tilde{t}}^4Y_q$ \\
\hspace{-3em}$+2M_{\tilde{q}}^2M_{\tilde{t}}^2(Y_q+Y_t)+M_{\tilde{q}}^4Y_t)\big)\cos2\beta\Big)\ln\frac{M_{\tilde{q}}^2}{M_{\tilde{t}}^2}$ \\
\hspace{-3em}$-(t\leftrightarrow b,g^2\to-g^2)$} \\
$+\frac{3y_t^4X_t^4M_{\tilde{q}}^2}{(M_{\tilde{q}}^2-M_{\tilde{t}}^2)^5}\Big(M_{\tilde{q}}^2+3M_{\tilde{t}}^2\Big)\ln\frac{M_{\tilde{q}}^2}{M_{\tilde{t}}^2}+(t\leftrightarrow b)$ \\
$+\frac{3y_t^2y_b^2X_t^2X_b^2}{(M_{\tilde{q}}^2-M_{\tilde{t}}^2)^4(M_{\tilde{q}}^2-M_{\tilde{b}}^2)^4(M_{\tilde{t}}^2-M_{\tilde{b}}^2)}$\specialcellt{$\Big[\Big((M_{\tilde{q}}^8-M_{\tilde{t}}^4M_{\tilde{b}}^4)$ \\
\hspace{-10.5em}$\times(M_{\tilde{t}}^4-M_{\tilde{b}}^4)-4M_{\tilde{q}}^2M_{\tilde{t}}^2M_{\tilde{b}}^2\big(M_{\tilde{t}}^2(M_{\tilde{q}}^4+M_{\tilde{b}}^4)$\\
\hspace{-10.5em}$-M_{\tilde{b}}^2(M_{\tilde{q}}^4+M_{\tilde{t}}^4)\big)\Big)\ln M_{\tilde{q}}^2$ \\
\hspace{-10.5em}$-(M_{\tilde{q}}^2-M_{\tilde{b}}^2)M_{\tilde{t}}^4\ln M_{\tilde{t}}^2$ \\
\hspace{-10.5em}$+(M_{\tilde{q}}^2-M_{\tilde{t}}^2)M_{\tilde{b}}^4\ln M_{\tilde{b}}^2\Big]$}
}
\\
\hline\hline
$(4\pi)^2c_{GG}$ & \specialcellc{$+\frac{1}{24M_{\tilde{q}}^2}\Big(y_t^2+y_b^2-g'^2Y_q\cos2\beta\Big)$ \\
$+\frac{1}{48M_{\tilde{t}}^2}\Big(2y_t^2+g'^2Y_t\cos2\beta\Big)+(t\leftrightarrow b)$ \\
$-\frac{y_t^2X_t^2}{24M_{\tilde{q}}^2M_{\tilde{t}}^2}-(t\leftrightarrow b)$}
& $0$ \\
\hline
$(4\pi)^2c_{WW}$ &
\specialcellc{$+\frac{1}{16M_{\tilde{q}}^2}\Big(y_t^2+y_b^2-g'^2Y_q\cos2\beta\Big)$
$+\frac{y_t^2X_t^2}{16M_{\tilde{q}}^2(M_{\tilde{q}}^2-M_{\tilde{t}}^2)}+(t\leftrightarrow b)$
}
& $-\frac{y_t^2X_t^2}{16(M_{\tilde{q}}^2-M_{\tilde{t}}^2)^2}\ln\frac{M_{\tilde{q}}^2}{M_{\tilde{t}}^2}-(t\leftrightarrow b)$ \\
\hline
$(4\pi)^2c_{BB}$ &
\specialcellc{$+\frac{Y_q^2}{4M_{\tilde{q}}^2}\Big(y_t^2+y_b^2-g'^2Y_q\cos2\beta\Big)$ \\
$+\frac{Y_t^2}{8M_{\tilde{t}}^2}\Big(2y_t^2+g'^2Y_t\cos2\beta\Big)+(t\leftrightarrow b)$ \\
$-\frac{y_t^2X_t^2}{4M_{\tilde{q}}^2M_{\tilde{t}}^2(M_{\tilde{q}}^2-M_{\tilde{t}}^2)}\Big(M_{\tilde{q}}^2Y_t^2-M_{\tilde{t}}^2Y_q^2\Big)-(t\leftrightarrow b)$
}
& $-\frac{y_t^2X_t^2(Y_q^2-Y_t^2)}{4(M_{\tilde{q}}^2-M_{\tilde{t}}^2)^2}\ln\frac{M_{\tilde{q}}^2}{M_{\tilde{t}}^2}-(t\leftrightarrow b)$ \\
\hline
$(4\pi)^2c_{WB}$ &
\specialcellc{$-\frac{Y_q}{8M_{\tilde{q}}^2}\Big(2y_t^2-2y_b^2+g^2\cos2\beta\Big)$ $-\frac{y_t^2X_t^2Y_q}{4M_{\tilde{q}}^2(M_{\tilde{q}}^2-M_{\tilde{t}}^2)}+(t\leftrightarrow b)$
}
& $+\frac{y_t^2X_t^2Y_q}{4(M_{\tilde{q}}^2-M_{\tilde{t}}^2)^2}\ln\frac{M_{\tilde{q}}^2}{M_{\tilde{t}}^2}-(t\leftrightarrow b)$ \\
\hline\newpage\hline
$(4\pi)^2c_W$ &
$-\frac{y_t^2X_t^2}{12(M_{\tilde{q}}^2-M_{\tilde{t}}^2)^4}\Big(5M_{\tilde{q}}^4-22M_{\tilde{q}}^2M_{\tilde{t}}^2+5M_{\tilde{t}}^4\Big)-(t\leftrightarrow b)$
& \specialcellc{$+\frac{y_t^2X_t^2}{4(M_{\tilde{q}}^2-M_{\tilde{t}}^2)^5}$\specialcellt{$\Big((M_{\tilde{q}}^2+M_{\tilde{t}}^2)(M_{\tilde{q}}^4-4M_{\tilde{q}}^2M_{\tilde{t}}^2+M_{\tilde{t}}^4)\Big)$ \\
\hspace{-3em}$\times\ln\frac{M_{\tilde{q}}^2}{M_{\tilde{t}}^2}+(t\leftrightarrow b)$}} \\
\hline
$(4\pi)^2c_B$ &
\specialcellc{$+\frac{y_t^2X_t^2}{36(M_{\tilde{q}}^2-M_{\tilde{t}}^2)^4}$\specialcellt{$\Big(\big(11+52(Y_q\hspace{-0.3em}-\hspace{-0.3em}Y_t)\big)(M_{\tilde{q}}^4+M_{\tilde{t}}^4)$ \\
\hspace{-3em}$+(19-28(Y_q\hspace{-0.3em}-\hspace{-0.3em}Y_t))2M_{\tilde{q}}^2M_{\tilde{t}}^2\Big)+(t\leftrightarrow b,Y\leftrightarrow-Y)$}}
& \specialcellc{$-\frac{y_t^2X_t^2}{12(M_{\tilde{q}}^2-M_{\tilde{t}}^2)^5}$\specialcellt{$\Big(9M_{\tilde{q}}^2M_{\tilde{t}}^2(M_{\tilde{q}}^2\hspace{-0.2em}+\hspace{-0.2em}M_{\tilde{t}}^2)
\hspace{-0.2em}+\hspace{-0.2em}\big(1+8(Y_q\hspace{-0.3em}-\hspace{-0.3em}Y_t)\big)$ \\
\hspace{-3em}$\times(M_{\tilde{q}}^6+M_{\tilde{t}}^6)\Big)\ln\frac{M_{\tilde{q}}^2}{M_{\tilde{t}}^2}+(t\leftrightarrow b,Y\leftrightarrow-Y)$}} \\
\hline
$(4\pi)^2c_{HW}$ &
$+\frac{y_t^2X_t^2}{2(M_{\tilde{q}}^2-M_{\tilde{t}}^2)^4}\Big(M_{\tilde{q}}^4-5M_{\tilde{q}}^2M_{\tilde{t}}^2-2M_{\tilde{t}}^4\Big)+(t\leftrightarrow b)$
& \specialcellc{$-\frac{y_t^2X_t^2}{4(M_{\tilde{q}}^2-M_{\tilde{t}}^2)^5}$\specialcellt{$\Big(M_{\tilde{q}}^6-3M_{\tilde{q}}^4M_{\tilde{q}}^2-9M_{\tilde{q}}^2M_{\tilde{t}}^4-M_{\tilde{t}}^6\Big)$ \\
\hspace{-3em}$\times\ln\frac{M_{\tilde{q}}^2}{M_{\tilde{t}}^2}-(t\leftrightarrow b)$}} \\
\hline
$(4\pi)^2c_{HB}$ &
\specialcellc{$-\frac{y_t^2X_t^2}{12(M_{\tilde{q}}^2-M_{\tilde{t}}^2)^4}$\specialcellt{$\Big((11+34Y_q+2Y_t)M_{\tilde{q}}^4+\big(19+8(Y_q\hspace{-0.3em}-\hspace{-0.3em}Y_t)\big)$  \\
\hspace{-3em}$\times2M_{\tilde{q}}^2M_{\tilde{q}}^2+(11-2Y_q-34Y_t)M_{\tilde{t}}^4\Big)+(t\leftrightarrow b,Y\leftrightarrow-Y)$} }
& \specialcellc{$+\frac{y_t^2X_t^2}{4(M_{\tilde{q}}^2-M_{\tilde{t}}^2)^5}$\specialcellt{$\Big((1+4Y_q)M_{\tilde{q}}^6+(3+4Y_q)3M_{\tilde{q}}^4M_{\tilde{t}}^2$ \\
\hspace{-3em}$+(3-4Y_t)3M_{\tilde{q}}^2M_{\tilde{t}}^4+(1-4Y_t)M_{\tilde{t}}^6\Big)\ln\frac{M_{\tilde{q}}^2}{M_{\tilde{t}}^2}$ \\
\hspace{-3em}$+(t\leftrightarrow b,Y\leftrightarrow-Y)$}
}\\
\hline
$(4\pi)^2c_D$ &
$\frac{y_t^2X_t^2}{2(M_{\tilde{q}}^2-M_{\tilde{t}}^2)^4}\Big(M_{\tilde{q}}^4+10M_{\tilde{q}}^2M_{\tilde{t}}^2+M_{\tilde{t}}^4\Big)+(t\leftrightarrow b)$
& $-\frac{3y_t^2X_t^2}{(M_{\tilde{q}}^2-M_{\tilde{t}}^2)^5}\Big(M_{\tilde{q}}^2M_{\tilde{t}}^2(M_{\tilde{q}}^2+M_{\tilde{t}}^2)\Big)\ln\frac{M_{\tilde{q}}^2}{M_{\tilde{t}}^2}-(t\leftrightarrow b)$ \\
\hline\hline
\label{tab:mainresult}
\end{longtable}

Here we have a few comments:
\begin{itemize}
\item In general every Wilson coefficient has a nonlogarithmic contribution and logarithmic contribution, the latter of which is proportional to the logarithm of the soft breaking mass ratio. This is the general way the scales nondegeneracy comes. In that case the denominators of both contributions contain factors of certain powers of the nondegeneracy, with the logarithmic one higher by one in power, consistent with the behavior of the Feynman parameter integration listed in the Appendix. It is straightforward to check that, while each contribution is singular in the degeneracy limit, the combination of the two are finite and reproducing an expansion around the degenerated results.
\item For a hierarchy of the large scales, only the nonlogarithmic contribution survives, the logarithmic contribution will always be more suppressed by one extra large scale square. It is straightforward to read them out, for example the leading contribution to $c_T$ is
\begin{align}
(4\pi)^2c_T\to
\left\{\begin{array}{lc}
{\displaystyle\frac{(2y_t^2-2y_b^2+g^2\cos2\beta)^2}{16M_{\tilde{q}}^2}+\frac{y_t^2(2y_t^2-2y_b^2+g^2\cos2\beta)X_t^2}{4M_{\tilde{q}}^2M_{\tilde{t}}^2}
+\frac{y_t^4X_t^4}{4M_{\tilde{q}}^2M_{\tilde{t}}^4}},\qquad & M_{\tilde{q}}\ll M_{\tilde{t}}, \\
{\displaystyle\frac{(2y_t^2-2y_b^2+g^2\cos2\beta)^2}{16M_{\tilde{q}}^2}-\frac{y_t^2(2y_t^2-2y_b^2+g^2\cos2\beta)X_t^2}{8M_{\tilde{q}}^4}
+\frac{y_t^4X_t^4}{4M_{\tilde{q}}^6}},\qquad & M_{\tilde{q}}\gg M_{\tilde{t}}.
\end{array}\right.
\end{align}
Note that in the second hierarchy even the leading contributions are suppressed by $M_{\tilde{q}}$.
\item The symmetry between the up type and the down type is manifested in the top-bottom switching $t\leftrightarrow b$, which by definition is $y_t\to y_b, M_{\tilde{t}}\to M_{\tilde{b}}, X_t\to X_b,Y_t\to Y_b$, and $y_b\to y_t, M_{\tilde{b}}\to M_{\tilde{t}}, X_b\to X_t,Y_b\to Y_t$ in some cases that the ``down'' type quantities also appear in a ``up'' type term. The $SU(2)_L$ couplings includes a $t^3$ induced flipping between the up and down sector, since they are the same for both squarks and sleptons we do not keep it explicitly like the hypercharge $Y_t(Y_b)$. Their effect is to flipping the sign of $g^2\to -g^2$. And note the sign flip of the $c_{WB}$, $c_B$ and $c_{HB}$ in the top-bottom switching.
\item In addition to the above operators, there are ``universal'' contributions to the pure gauge boson operators. Since the new particle couplings to SM gauge group are always within a gauge representation so that share the same large scale, for example in our sfermion case the $SU(2)_L$ gauge bosons only couple to left hand sfermions so that only feel the $M_{\tilde{q}}$($M_{\tilde{l}}$), and the other $SU(3)_c$ and $U(1)_Y$ gauge bosons couplings are block diagonal in the $M^2+\delta V''$ matrix, they do not show the above nondegeneracy property. Here we adapt the results in~\cite{Henning:2014wua} for completeness
\begin{equation}
\mathcal{L}_\text{EFT}\supset\frac{1}{(4\pi)^2}\bigg(\Big(\frac{Y_q^2g'^2}{10M_{\tilde{q}}^2}+\frac{Y_t^2g'^2}{20M_{\tilde{t}}^2}+\frac{Y_b^2g'^2}{20M_{\tilde{b}}^2}\Big){\cal O}_{2B}+\frac{g^2}{20M_{\tilde{q}}^2}\Big({\cal O}_{2W}+{\cal O}_{3W}\Big)+ \Big(\frac{g_s^2}{30M_{\tilde{q}}^2}+\frac{g_s^2}{60M_{\tilde{t}}^2}+\frac{g_s^2}{60M_{\tilde{b}}^2}\Big)\Big({\cal O}_{2G}+{\cal O}_{3G}\Big)\bigg).
\label{eq:hitoshi}
\end{equation}
Only the operator ${\cal O}_{3W}$ is directly constrained by the tri-gauge boson precision experiments, and is generally expected to have a relatively low sensitivity. We do not include them in the following fit.
\item After the original version of this paper, \cite{Drozd:2015rsp} appeared and pointed out some discrepancies between the Wilson coefficients here and theirs. In this updated version the omission in the first calculation has been corrected, which matches with~\cite{Drozd:2015rsp}. The calculation here differs from~\cite{Drozd:2015rsp} in the trick of expanding the logarithm in the effective potential, see~\cite{Han:2017cfr} for discussion of the regularization options. The degenerated mass case result is obtained in~\cite{Henning:2014gca,Henning:2014wua}. The nondegenerate mass results for $c_{GG}$, $c_{WW}$, $c_{BB}$ and $c_{WB}$ are obtained in~\cite{Drozd:2015kva}, however the latter three are in a different basis. Here it is straightforward to check that for the Higgs diphoton coupling the combination of terms proportional to $y_t^2X_t^2$ from the latter three operators conspires to cancel the $(M_{\tilde{q}}^2-M_{\tilde{t}}^2)^{-1}$ and the logarithmic term, going back to a form consistent with the Higgs low energy theorem. The alternative Feynman diagram calculation dates back to~\cite{Drees:1991zk}.
\item The results which focus at one loop level and dimension-6 operators here, are the leading contributions to various SM precision test of integrating out the MSSM sfermion sector. Further improvement can be made by going beyond one loop level or operators of dimension 6 or both. The former improvement includes not only a straightforward two loop calculation, but also other effects such as an renormalization group running of the one loop sized contributions from the matching scale $M_{\tilde{t}}$ to the individual scales for each precision experiment~\cite{Masso:2012eq}, or the effort to make the definition of $T$ and $S$ parameters gauge invariant~\cite{Chen:2013kfa}. The latter improvement on the other hand, is recovered in the traditional Feynman diagram calculation, in which the one loop results are not truncated at dimension-6 operators level. But according to the comparison made in~\cite{Drozd:2015kva}, for our interested $M_{\tilde{t}_1}>500~\text{GeV}$ region the difference between the two methods are negligible.
\item In the following numerical works we will focus on the four most stringent constraints, but completely matching to EFT indeed has an advantage of providing much more information. For example, following~\cite{Henning:2014wua} the $hZ\gamma$, $hWW$, $hZZ$ can also be calculated and constrained with the Wilson coefficients.
\end{itemize}

\section{Numerical Constraints \label{sec:num}}

The afore calculated Wilson coefficients can be used straightforwardly to transfer the model independent constraints to a set of constraints of the sfermion parameters. In the following we will ignore the subdominant slepton sector contribution, and in the squark sector we will ignore the much smaller bottom Yukawa couplings. Consequently the dependence on $X_b$ is eliminated, for $X_b$ is always multiplied by $y_b$\footnote{In the large $\tan\beta$ region this can be not true, but for simplicity here we will not consider such scenario.}. In that case the $M_{\tilde{b}}$ dependence is always proportional to only the small $g'$ and therefore weak, for simplicity we assume $M_{\tilde{b}}=M_{\tilde{t}}$\footnote{Numerically if $M_{\tilde{b}}$ varies by a factor of two, for the following typical parameter choices the most sensitive $c_{GG}$ will vary by a few percent.}. While the results in Table~\ref{tab:mainresult} depend on the two soft breaking parameters $M_{\tilde{q}}$ and $M_{\tilde{t}}$, they can be translated to the physical masses of the two stop squarks, through mixing determined by diagonalization of stop mixing matrix
\begin{equation}
M^2+\delta V''=\left(\begin{array}{cc}
M_{\tilde{q}}^2+m_t^2+m_Z^2(\frac{1}{2}-\frac{2}{3}\sin^2\theta_W)\cos2\beta & m_tX_t  \\
m_tX_t & M_{\tilde{t}}^2+m_t^2+m_Z^2\frac{2}{3}\sin^2\theta_W\cos2\beta
\end{array}\right).\label{eq:stop}
\end{equation}
Note that the mixing matrix is just the first and third rows and columns of the matrix $V''=M^2+\delta V''_F+\delta V''_D+\delta V''_X$ in Eqs.~(\ref{eq:M^2},\ref{eq:VF},\ref{eq:VD},\ref{eq:VX}), while the Higgs components are replaced by their VEVs. We will ignore any loop correction to the above squark mass matrix.

There are four most stringent operator level constraints in the SM EFT. Two of them are the EWPT Peskin-Takeuchi $T$ and $S$ parameters~\cite{Peskin:1990zt}, and the other two correspond to two channels of Higgs production/decay coupled to gluon pairs and photon pairs, which are generated at one-loop level in the SM even in leading order. The four constraints in our operator basis of Table~\ref{tab:Op} are represented~\cite{Henning:2014wua}
\begin{align}
T=&\frac{1}{\alpha_\text{EM}}v^2c_T,\quad&\quad S=&4\pi v^2(4c_{WB}+c_W+c_B),\\
\frac{\Delta\Gamma_{hgg}}{\Gamma^\text{SM}_{hgg}}=&\frac{(4\pi)^2}{\text{Re}\mathcal{M}_{hgg}}8v^2c_{GG},\quad&\quad \frac{\Delta\Gamma_{h\gamma\gamma}}{\Gamma^\text{SM}_{h\gamma\gamma}}=&\frac{(4\pi)^2}{\text{Re}\mathcal{M}_{h\gamma\gamma}}4v^2(c_{WW}+c_{BB}-c_{WB}),
\end{align}
where $\mathcal{M}_{hgg}$ and $\mathcal{M}_{h\gamma\gamma}$ is the SM amplitude of the (looped) $hgg$ and $h\gamma\gamma$ couplings. Note in the Higgs precision experiments usually measurements are not for $\frac{\Delta\Gamma_{hgg}}{\Gamma^\text{SM}_{hgg}}$($\frac{\Delta\Gamma(h\gamma\gamma)}{\Gamma_\text{SM}(h\gamma\gamma)}$) or $\frac{\Delta \text{BR}}{\text{BR}}$ but for $\frac{\Delta \sigma\cdot\text{BR}}{\sigma\cdot\text{BR}}$, and there are subdominant contributions induced by other operators beyond the interference of new physics and the SM amplitude, but for simplicity we will ignore all of them. The current best measurements and projected future sensitivities are listed in Table~\ref{tab:sigs}.
\begin{table}[th]
\begin{tabular}{ccccc}
\hline\hline
Observable & ~~~~~~$T$~~~~~~ & ~~~~~~$S$~~~~~~ & ~~~~~~$\Delta\Gamma_{hgg}/\Gamma^\text{SM}_{hgg}$~~~~~~ & ~~~~~~$\Delta\Gamma_{h\gamma\gamma}/\Gamma^\text{SM}_{h\gamma\gamma}$~~~~~~  \\
\hline
Current & $0.07$~\cite{Baak:2014ora} & $0.09$~\cite{Baak:2014ora} & $4.6\%$~\cite{Ellis:2014jta} &  \\
ILC (1TeV) & $0.022$~\cite{Baak:2014ora} & $0.016$~\cite{Baak:2014ora} & $3.1\%$~\cite{Baer:2013cma} & $8.5\%$~\cite{Baer:2013cma} \\
CEPC & $0.009$~\cite{Fan:2014vta} & $0.014$~\cite{Fan:2014vta} & $1.9\%$~\cite{Fan:2014vta} & $9.1\%$~\cite{Fan:2014vta} \\
FCC-ee & $0.004$~\cite{TeraZ} & $0.007$~\cite{TeraZ} & $1.4\%$~\cite{Gomez-Ceballos:2013zzn} & $3.0\%$~\cite{Gomez-Ceballos:2013zzn} \\
\hline\hline
\end{tabular}
\caption{The uncertainties achieved or expected at each experiments, for the four most constraining model independent operators or observables. }
\label{tab:sigs}
\end{table}

\begin{figure}[th]
\centering
\includegraphics[height=2in]{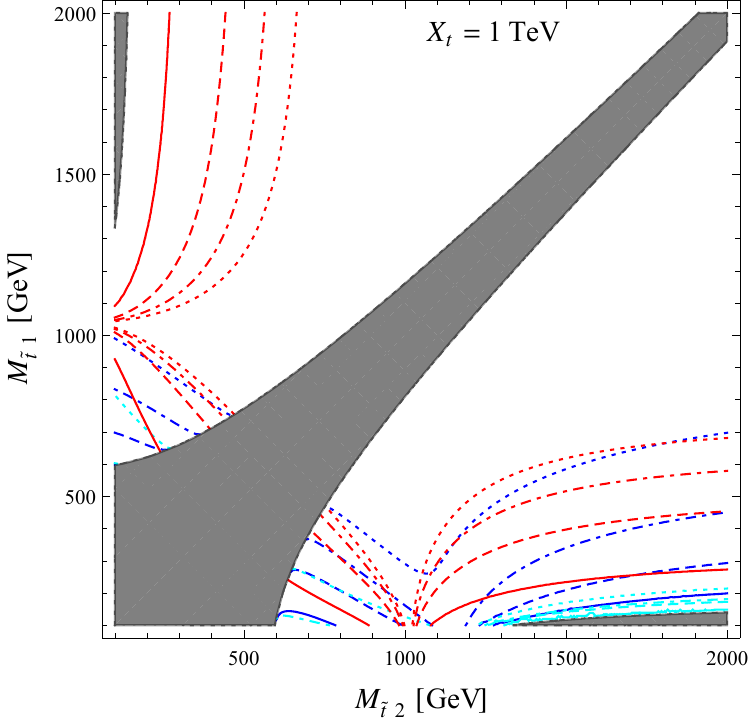}\quad
\includegraphics[height=2in]{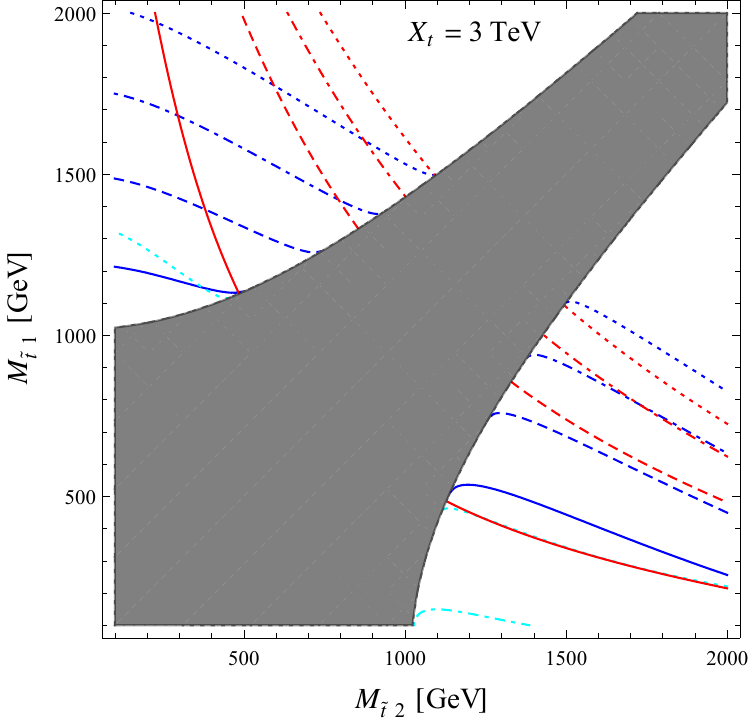}\quad
\includegraphics[height=2in]{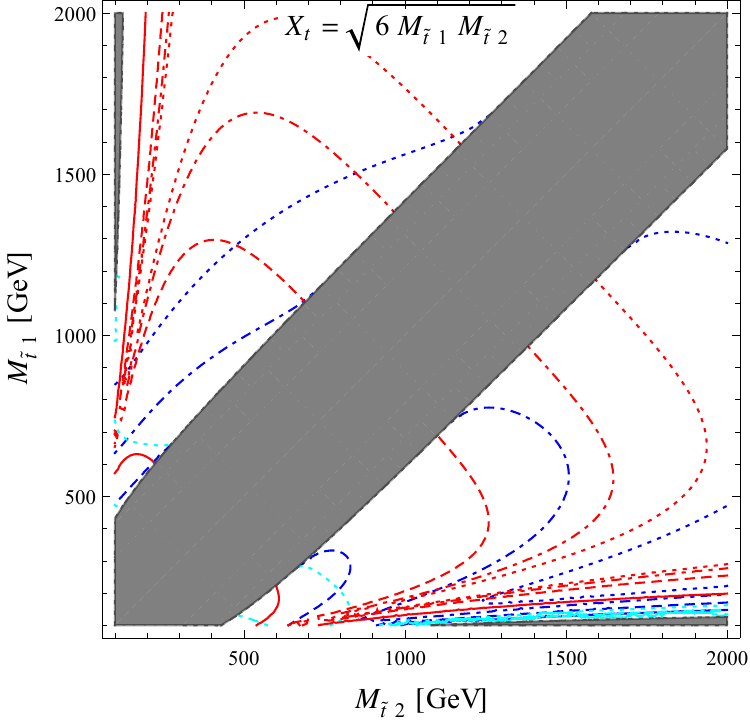}
\caption{The $2\sigma$ $T$ parameter (blue curves), $S$ parameter (cyan curves), and $hgg$ coupling (red curves) constraints, shown for the left $X_t=1~\text{TeV}$, middle $X_t=3~\text{TeV}$ and right $X_t=\sqrt{6M_{\tilde{t}_1}M_{\tilde{t}_2}}$ panels. For each group of the $T$, $S$, and $hgg$ curves the current constraints, ILC expected constraints, CEPC expected constraints and FCC-ee expected constraints are denoted by solid curves, dashed curves, dot-dashed curves, and dotted curves respectively, if the constraint is strong enough to be shown on the plot. The shaded region is theoretically inaccessible for the $X_t$ choice with mixing indicated by Eq.~(\ref{eq:stop}). Here we choose $\tan\beta=20$ but the $\tan\beta$ dependence is weak.}
\label{fig:fixXt}
\end{figure}

A lot of numerical fittings have already been performed in the literature~\cite{Craig:2014una,Fan:2014axa,Drozd:2015kva}. In Fig.~\ref{fig:fixXt} we focus on the $2\sigma$ constraints of the $T$ and $S$ operators as well as the $hgg$ coupling for comparison with~\cite{Drozd:2015kva} for different choices of $X_t$, with nondegenerate $M_{\tilde{q}}$ and $M_{\tilde{t}}$ substituted by the physical stop masses $M_{\tilde{t}_1}$ and $M_{\tilde{t}_2}$. The $X_t$ introduces some splitting between the two stop masses, so for nonzero $X_t$ a region with $M_{\tilde{t}_1}\simeq M_{\tilde{t}_2}$ is inaccessible and shaded gray in the plots (also regions with very small $M_{\tilde{t}_1}$ and large $M_{\tilde{t}_2}$). We assume $\tilde{t}_1$ has larger mixing component of left handed stop and $\tilde{t}_2$ of right handed stop, so that the bottom right corner of each plot shows the constraints for $M_{\tilde{q}}<M_{\tilde{t}}$, while the $M_{\tilde{q}}>M_{\tilde{t}}$ case is shown in the top left corner. Since the $X_t=0$ slice will not satisfy the SM like Higgs mass constraint of $122~\text{GeV}<M_h<128~\text{GeV}$ in the MSSM anyway (except for exponentially large soft mass) we will not show such result, and the three plots shown are for $X_t=1$~TeV, $3$~TeV, and $\sqrt{6M_{\tilde{t}_1}M_{\tilde{t}_2}}$ respectively.

In all the three panels the most constraining operator arises between the $hgg$ and $T$. The $hgg$ coupling constraints are usually the most stringent, with exceptions on the $M_{\tilde{q}}<M_{\tilde{t}}$ side when the two stops mass splitting is large, in that case the $T$ parameter constraints will take over. In the first panel the constraints vanish at $M_{\tilde{t}_1}\simeq X_t$ or $M_{\tilde{t}_2}\simeq X_t$ of $1~\text{TeV}$, that's the ``blind spot'' discussed in~\cite{Craig:2014una,Fan:2014axa}. Except for the blind spot, the constraint for the heavier stops will naively extend to large values, that's another interesting feature of the indirect constraint method. We have checked that the constraints depend very weakly with $\tan\beta$, which can be understood through the fact that it always comes in the $\cos2\beta$ factor of the small D-term.

In supersymmetric theories the SM like Higgs mass is calculable, from the D-term tree level contribution and the large radiative corrections mainly from the top/stop sector. For stop masses of a few TeVs, the $X_t$ term needs to contribute significantly to tune the Higgs mass to the measured $125$~GeV, which predicts $X_t$ itself to be a considerable value with roughly the same scaling with stop masses, potentially contributing significantly through the $X_t^2$ and $X_t^4$ terms in the Wilson coefficients. So the SM like Higgs mass imposes another constraint of the parameter region, and it is interesting to stick to that parameter slice. Note that there is still some uncertainty in calculating the SM like Higgs mass, especially in the low SUSY scale region, different code may give up to $3$~GeV discrepancy. We use the \texttt{FeynHiggs 2.11.2}~\cite{Heinemeyer:1998yj} to calculate the MSSM Higgs mass (for an estimation of hierarchical stops two-loop contribution, see Eq. (5.3) of~\cite{Batell:2013psa}), then solve for the $X_t$ values which are consistent with a SM like Higgs mass of $122\sim128~\text{GeV}$.

\begin{figure}[th]
\centering
\includegraphics[height=2in]{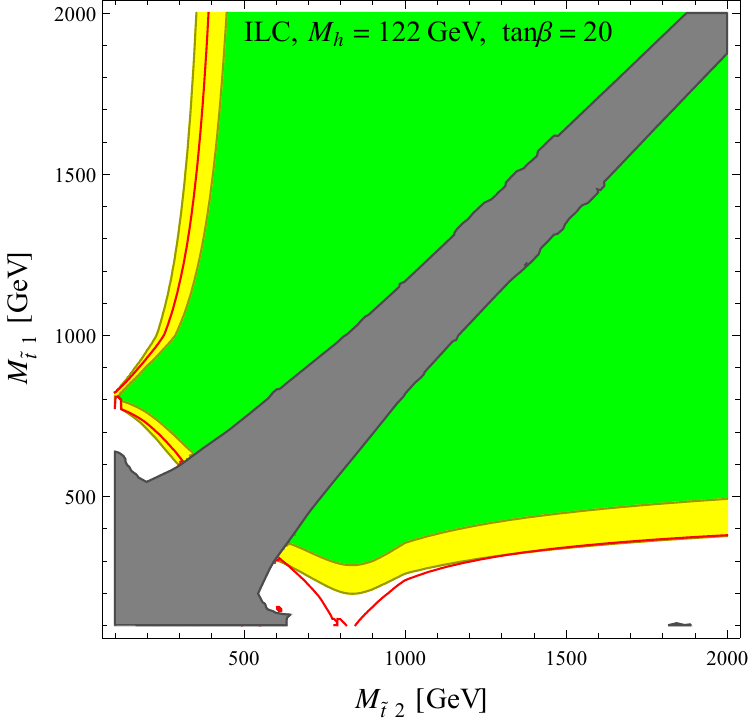}\quad
\includegraphics[height=2in]{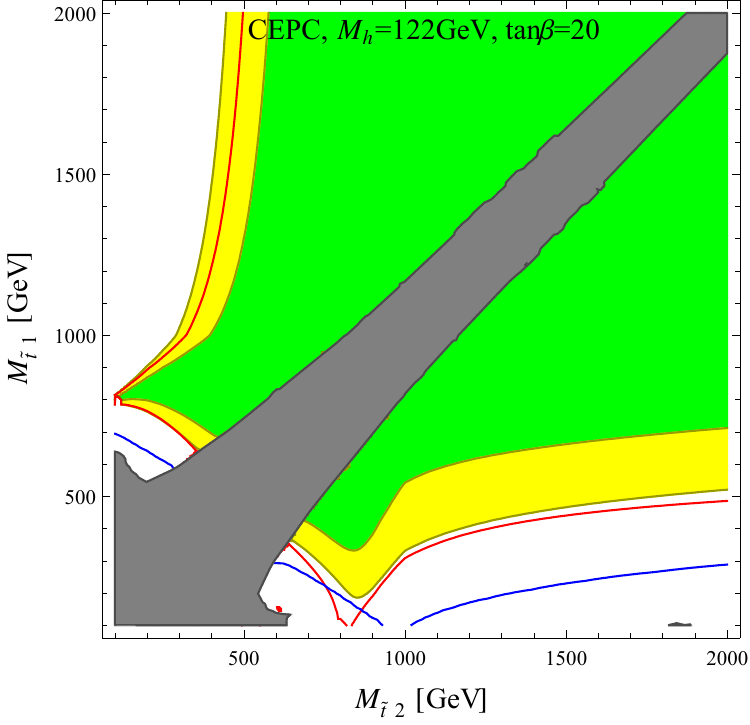}\quad
\includegraphics[height=2in]{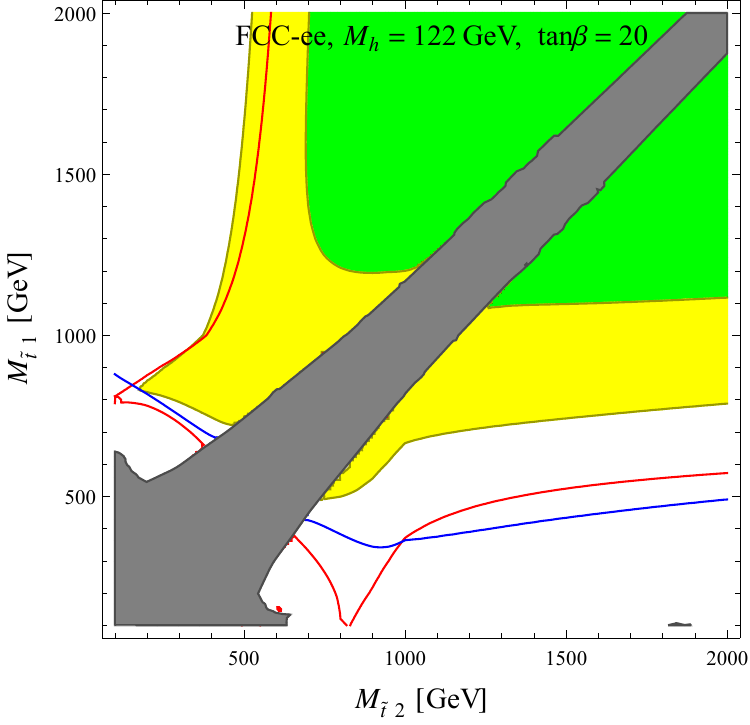}\\
\includegraphics[height=2in]{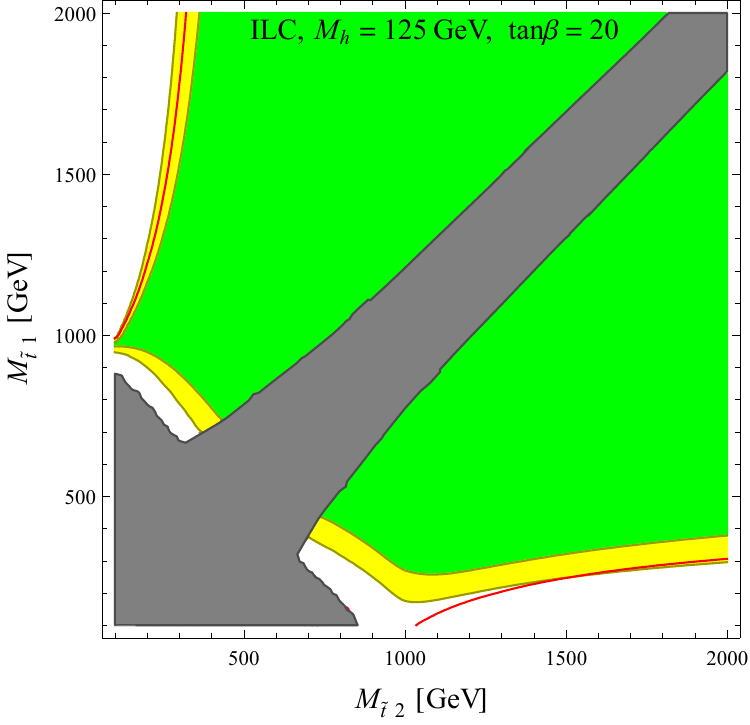}\quad
\includegraphics[height=2in]{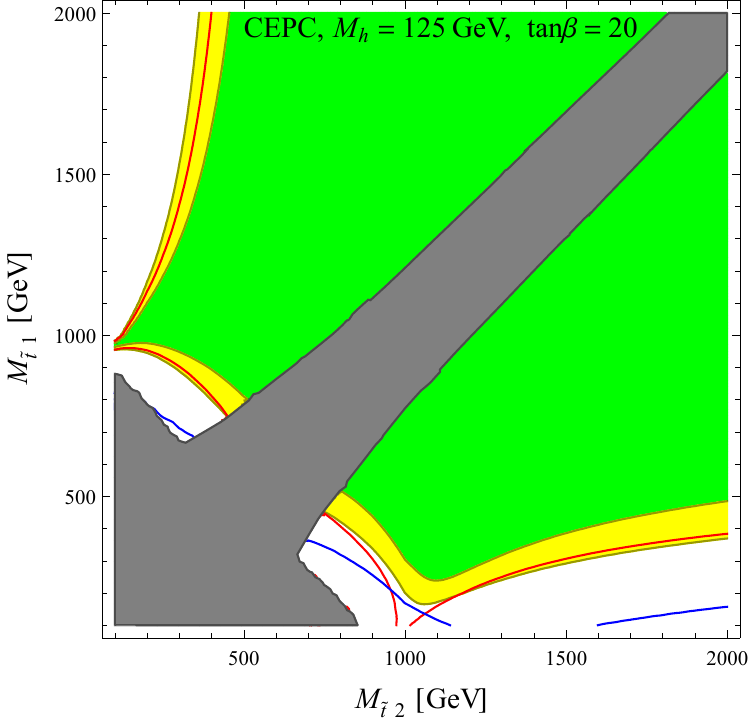}\quad
\includegraphics[height=2in]{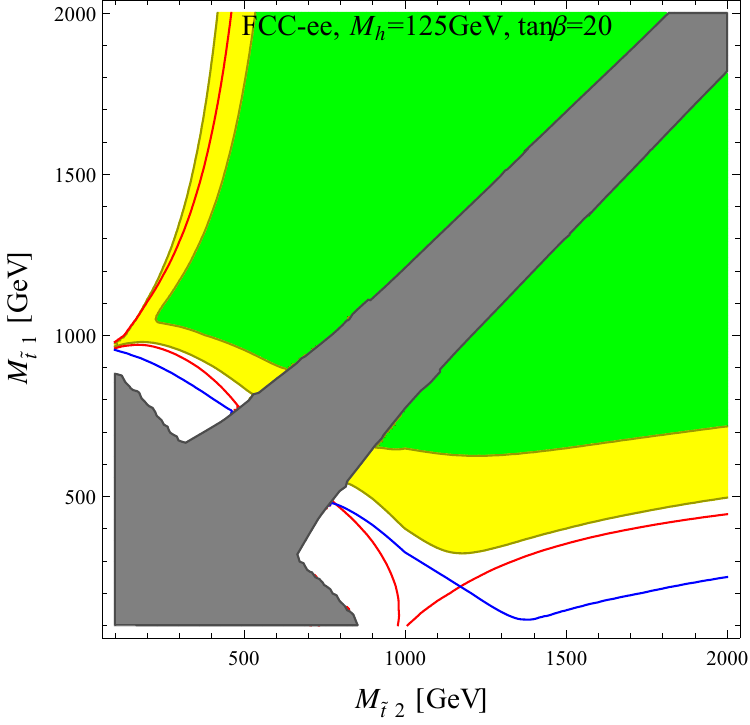}\\
\includegraphics[height=2in]{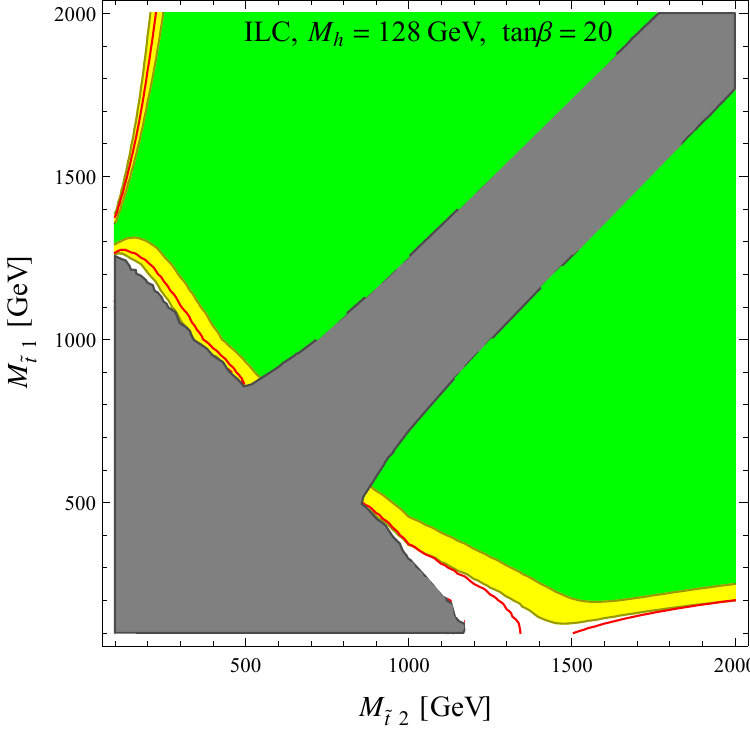}\quad
\includegraphics[height=2in]{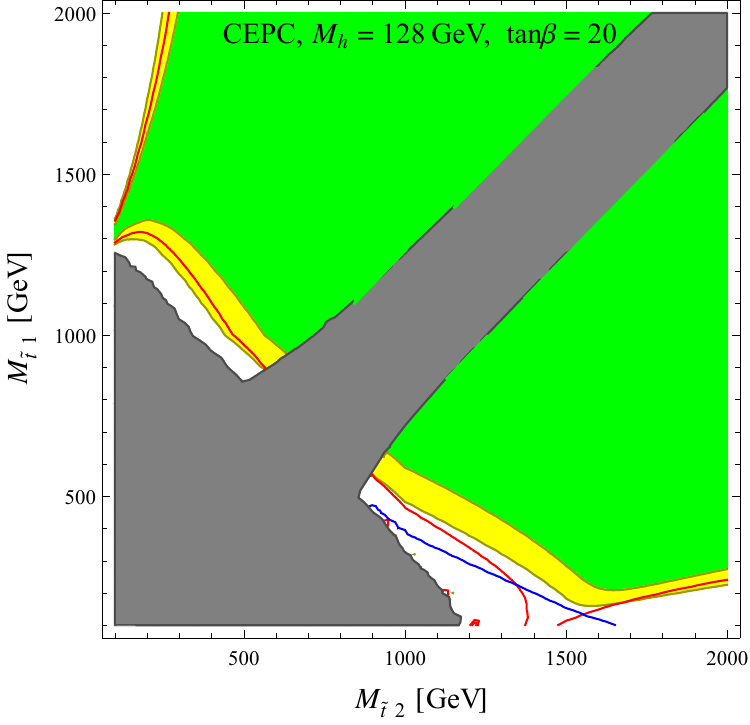}\quad
\includegraphics[height=2in]{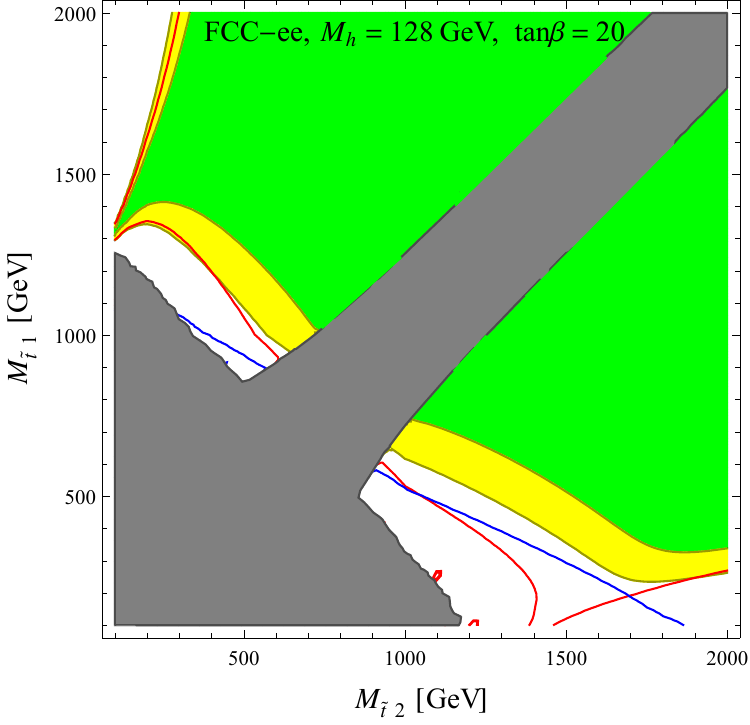}
\caption{The expected constraints from future ILC (the left column), CEPC (the middle column), and FCC-ee (the right column) experiments (see Table~\ref{tab:sigs}) for fixed SM like Higgs mass value of $M_h=122~\text{GeV}$ (the first row), $M_h=125~\text{GeV}$ (the second row), and $M_h=128~\text{GeV}$ (the third row). The $1\sigma$ and $2\sigma$ allowed region are green and yellow hatched. The individual $2\sigma$ constraint from $T$, $S$ parameters and $hgg$, $h\gamma\gamma$ couplings are shown for each experiment again as blue, cyan, red, and magenta curves respectively, if strong enough to be shown. There are still theoretically inaccessible region as hatched in gray. Here we choose $\tan\beta=20$, and the small $X_t$ solution to reproduce the required SM like Higgs mass.}
\label{fig:fixXt120}
\end{figure}

\begin{figure}[th]
\centering
\includegraphics[height=2in]{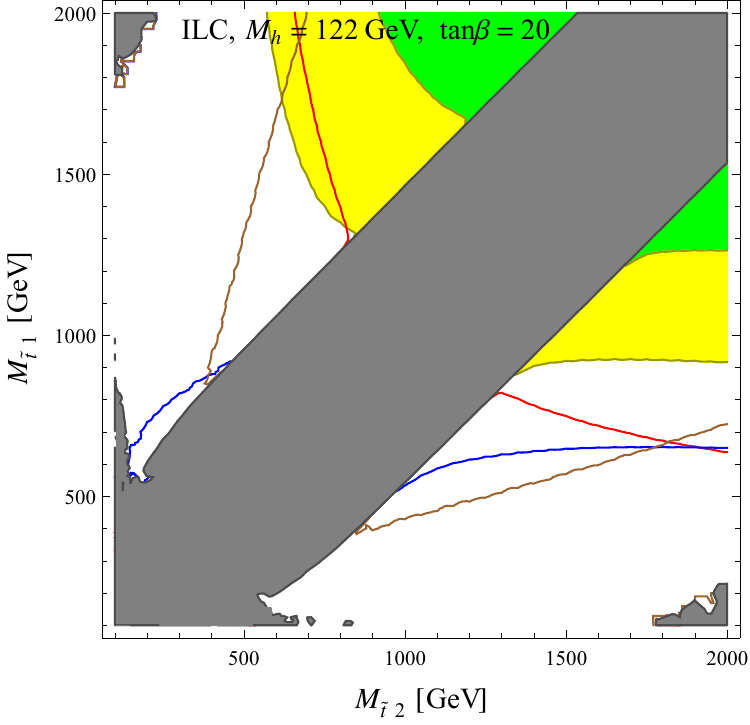}\quad
\includegraphics[height=2in]{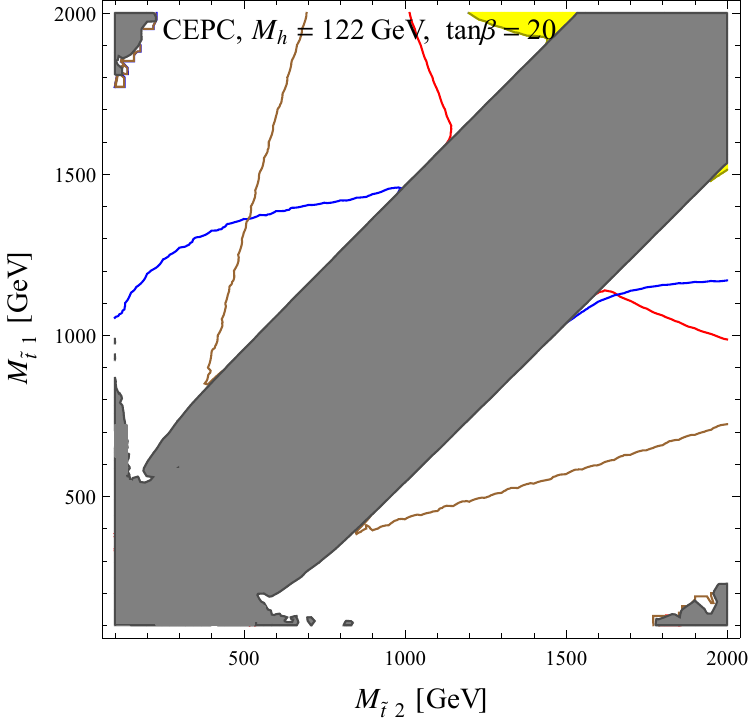}\quad
\includegraphics[height=2in]{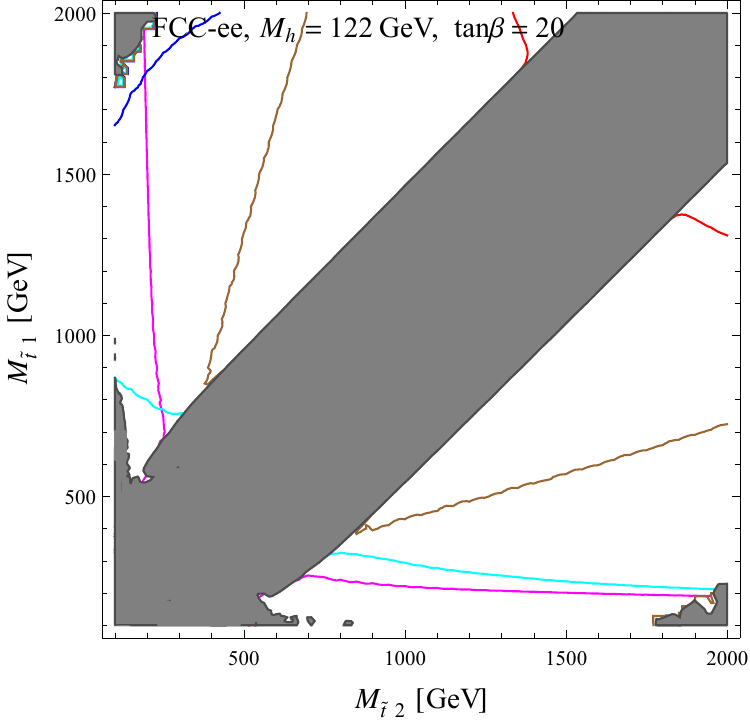}\\
\includegraphics[height=2in]{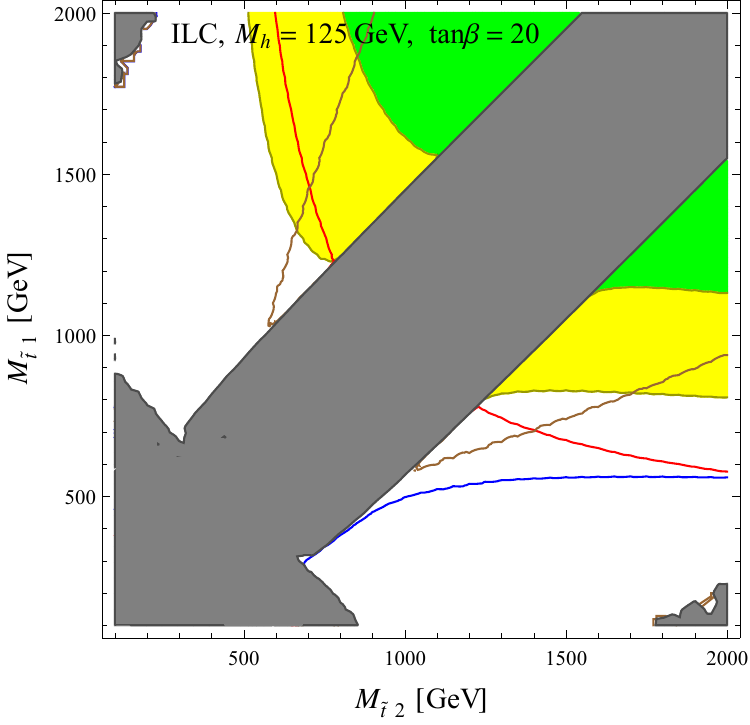}\quad
\includegraphics[height=2in]{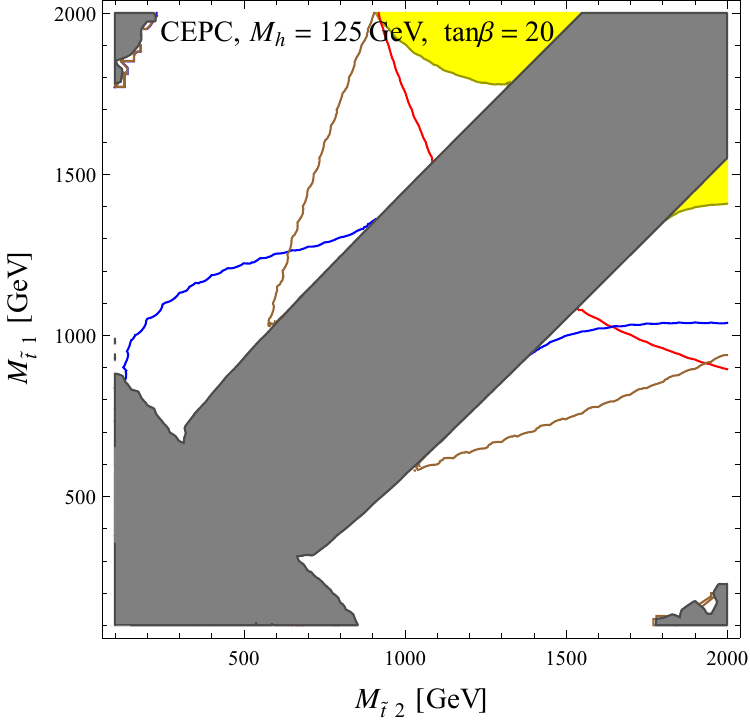}\quad
\includegraphics[height=2in]{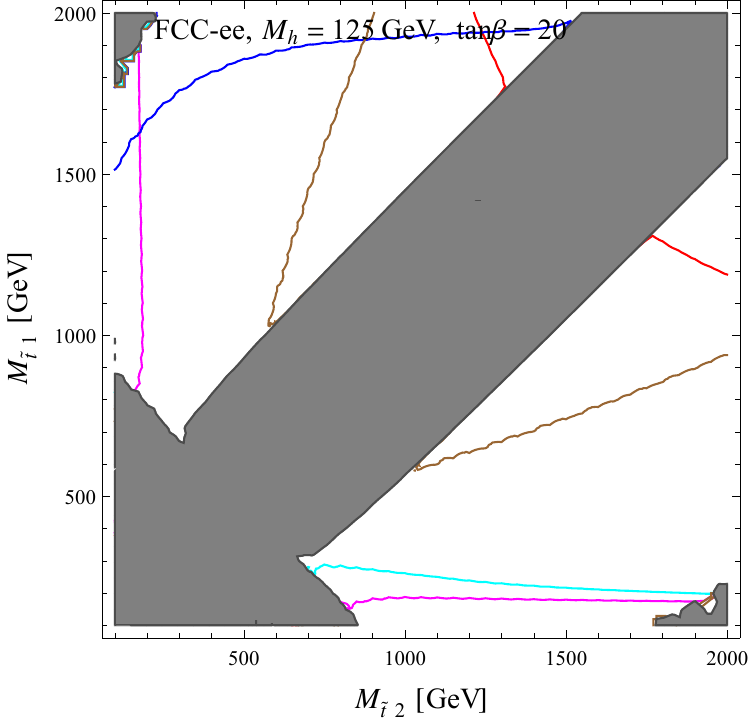}\\
\includegraphics[height=2in]{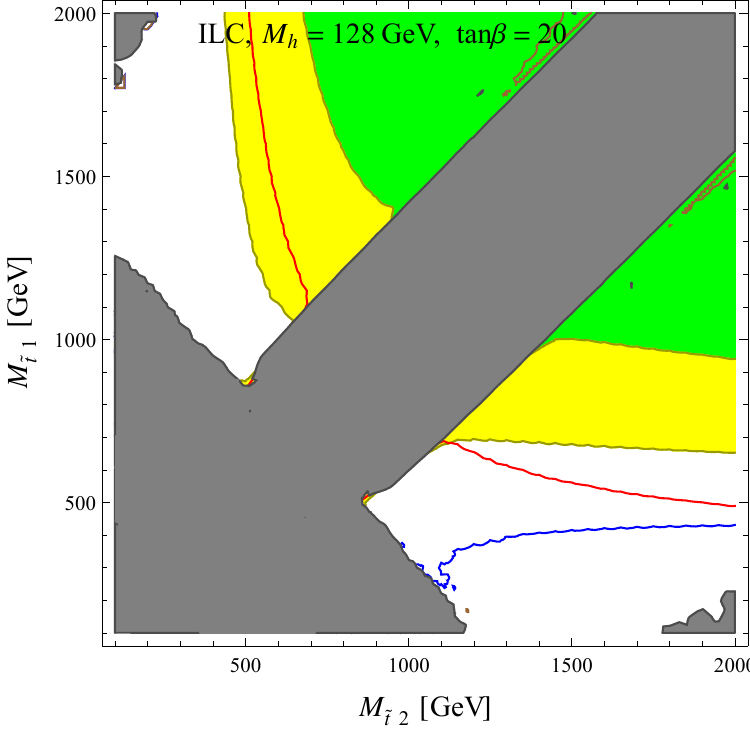}\quad
\includegraphics[height=2in]{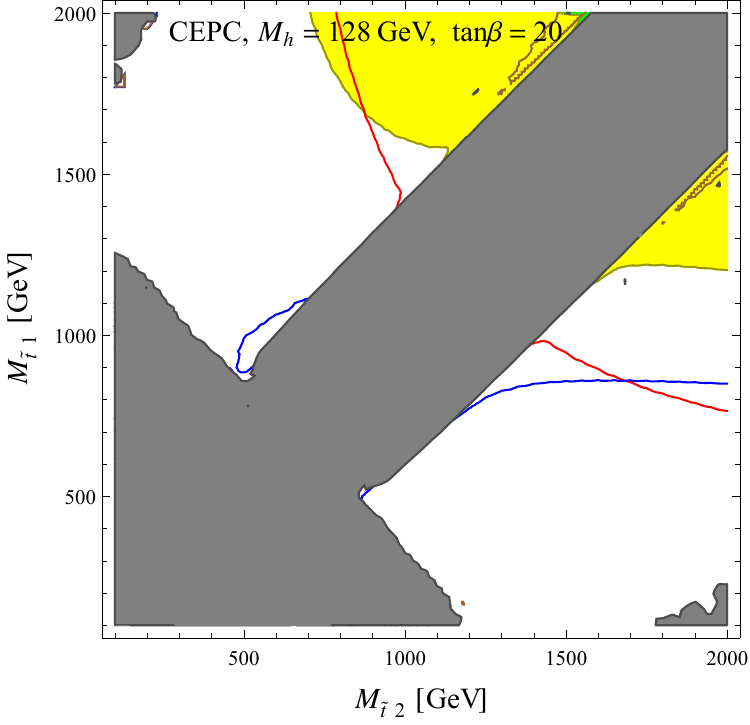}\quad
\includegraphics[height=2in]{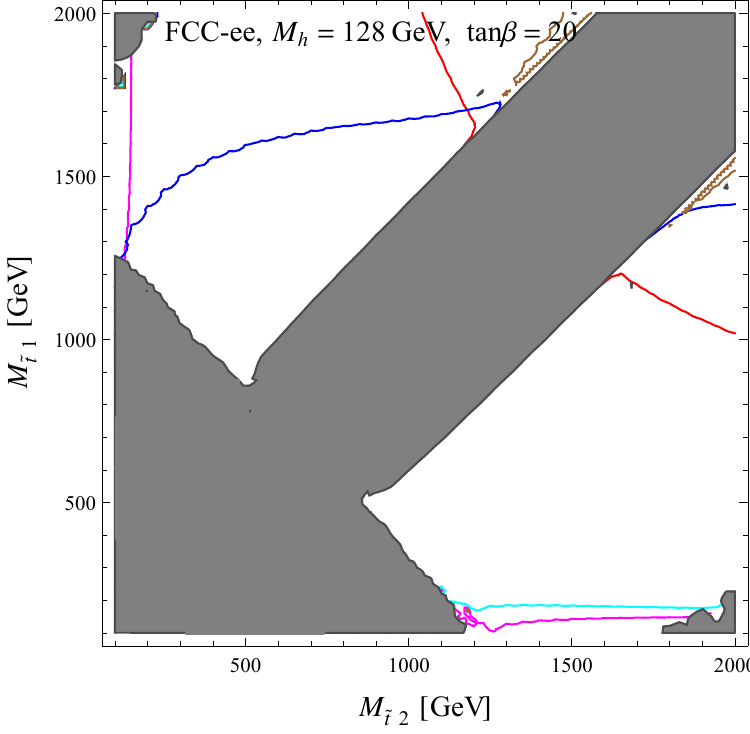}
\caption{Same as Fig.~\ref{fig:fixXt120}, but for large $X_t$ solution to reproduce the required SM like Higgs mass. The brown curves are the charge and color breaking vacuum constraint of $X_t/\sqrt{M_{\tilde{t}_1}^2+M_{\tilde{t}_2}^2}\lesssim\sqrt{3}$, and the region below and on the left are allowed.}
\label{fig:fixXt220}
\end{figure}

In Figs.~\ref{fig:fixXt120} and \ref{fig:fixXt220} we show such numerical constraints for three future experiments and a range of SM like Higgs masses at $\tan\beta=20$ and decoupling large CP odd Higgs mass. Again we use the same notations and methods for the two cases of $M_{\tilde{q}}<M_{\tilde{t}}$ and $M_{\tilde{q}}>M_{\tilde{t}}$, and gray shading for theoretically inaccessible region. The green $1\sigma$ region and yellow $2\sigma$ region are determined by fitting to all the EWPT and Higgs precision observables in~\cite{Baak:2014ora,Baer:2013cma,Fan:2014vta,TeraZ,Gomez-Ceballos:2013zzn} (including the $hWW$, $hZZ$, and Higgs to fermion pairs expectations ignored in Table~\ref{tab:sigs}), and four individual $2\sigma$ constraints corresponding to Table~\ref{tab:sigs} are shown as well. For a fixed SM like Higgs mass value there are in general two $X_t$ solutions as shown respectively in the two sets of plots, and the most effective constraints are seen in the large $X_t$ branch. While the CEPC can probe most low stop mass ($<2~\text{TeV}$) region except for two corners, the FCC-ee can completely cover the whole region, even with a $3~\text{GeV}$ SM like Higgs mass uncertainty. On the large $X_t$ branch one also need to worry about the possibility of other vacuum deeper than our electroweak symmetry breaking one with nonzero VEV of stops scalars, which may be induced by the large nondiagonal $X_t$. We show the bound of $X_t/\sqrt{M_{\tilde{t}_1}^2+M_{\tilde{t}_2}^2}\lesssim\sqrt{3}$ of~\cite{Camargo-Molina:2013sta,Chowdhury:2013dka,Blinov:2013fta} in Fig.~\ref{fig:fixXt220} for comparison.

\begin{figure}[th]
\centering
\includegraphics[height=2in]{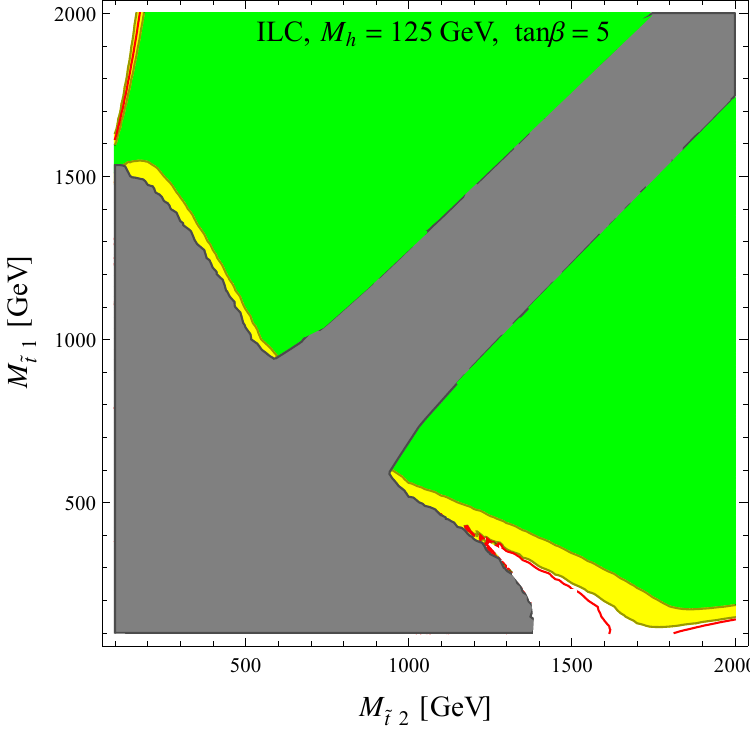}\quad
\includegraphics[height=2in]{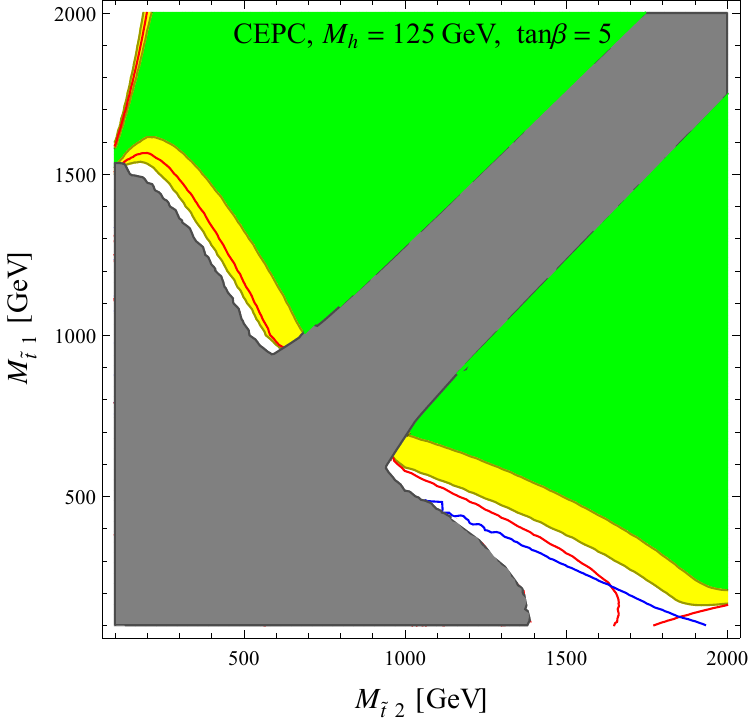}\quad
\includegraphics[height=2in]{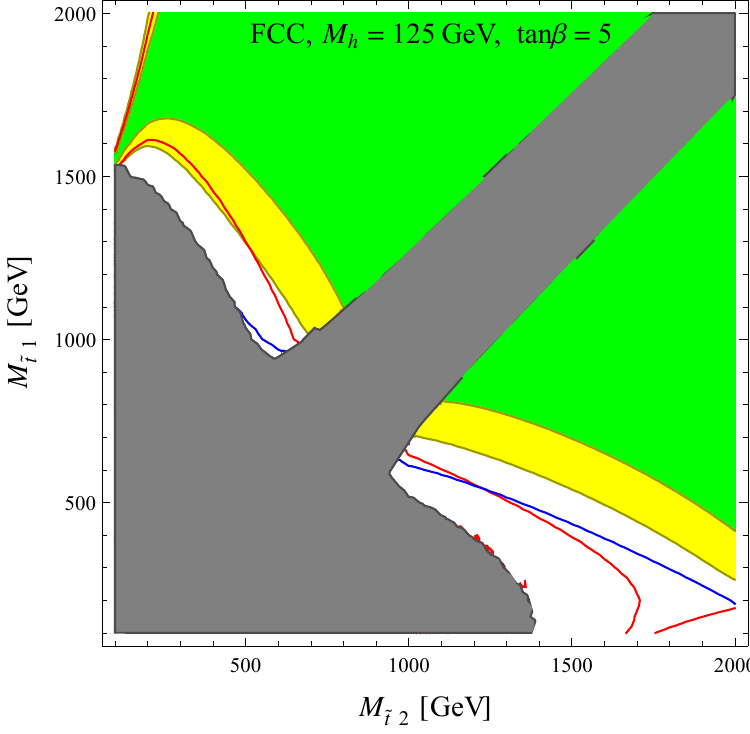}\\
\includegraphics[height=2in]{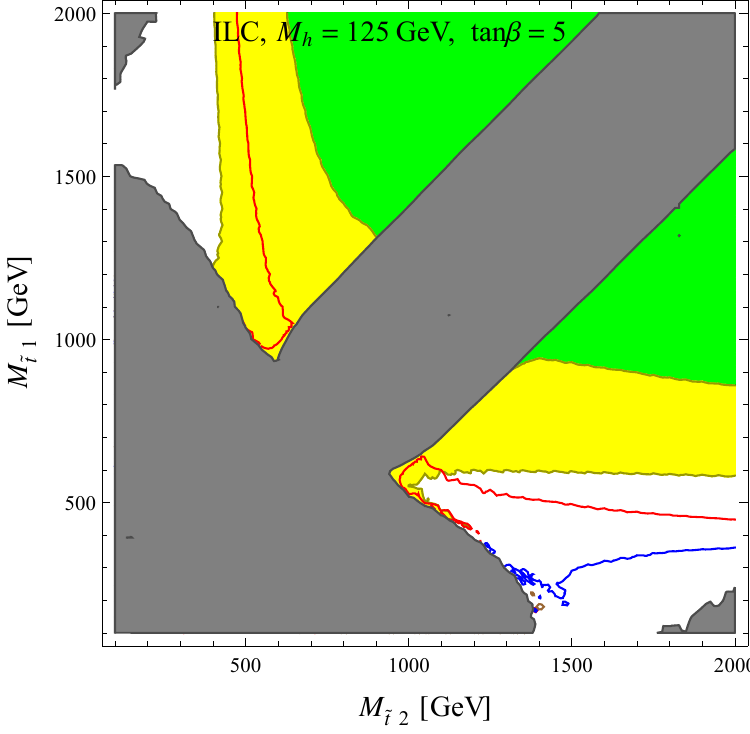}\quad
\includegraphics[height=2in]{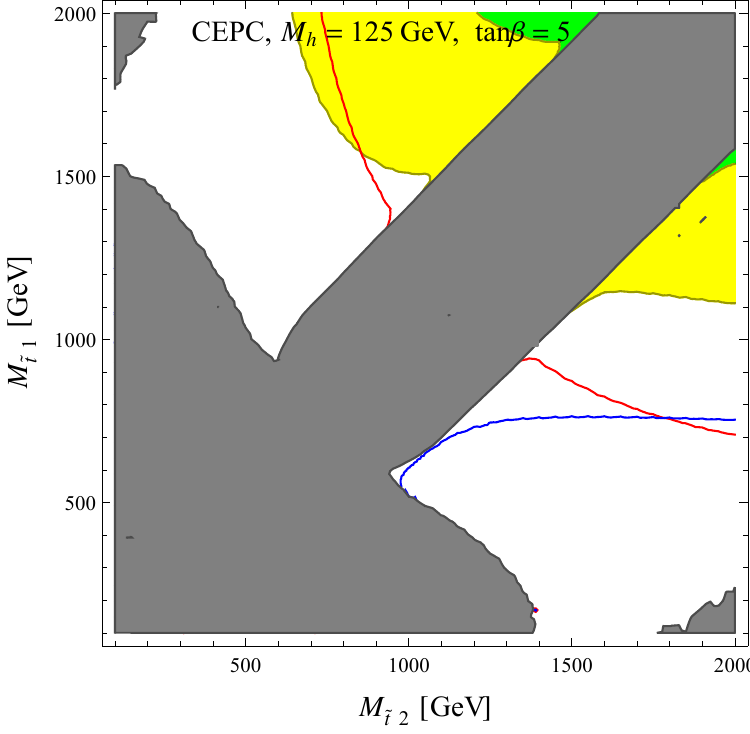}\quad
\includegraphics[height=2in]{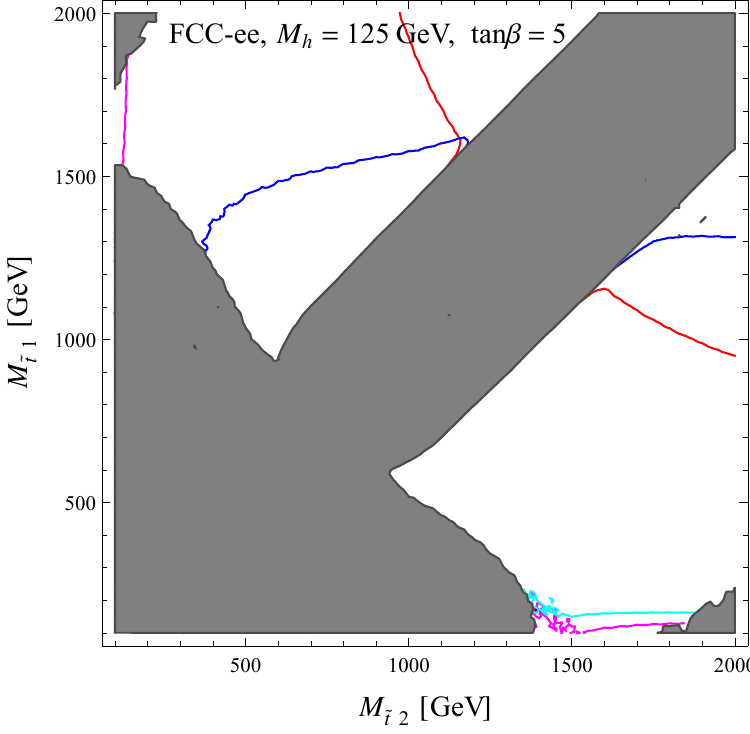}
\caption{The expected constraints from future ILC (the left column), CEPC (the middle column) and FCC-ee (the right column) experiments (see Table~\ref{tab:sigs}) for fixed SM like Higgs mass value of $M_h=125~\text{GeV}$, with small $X_t$ solution (the first row) and large $X_t$ solution (the second row). Other notations are the same with Fig.~\ref{fig:fixXt120}. In the large $X_t$ branch all the theoretically accessible region are consistent with the charge and color breaking vacuum constraint of $X_t/\sqrt{M_{\tilde{t}_1}^2+M_{\tilde{t}_2}^2}\lesssim\sqrt{3}$.}
\label{fig:fixXt5}
\end{figure}

At last we check the $\tan\beta$ dependence, which has a large effect on the SM like Higgs mass. As can be seen from Fig.~\ref{fig:fixXt5}, smaller $\tan\beta$ will allow a smaller $X_t$ on the large $X_t$ branch, so that weaken the constraints compared with the large $\tan\beta$ case. Still we can see that even if $\tan\beta$ is as low as $5$, the low stop mass ($<2~\text{TeV}$) region on the large $X_t$ branch can be almost completely probed.

\section{Conclusion \label{sec:dis}}

In this paper as an example of the CDE generalization, we perform the one-loop integration out of the sfermion sector in the MSSM, with full respect to all the coupling constants and nondegeneracy of the soft mass squares, and matching it to a basis of dimension-6 pure bosonic operators. Analytical expressions are given for each Wilson coefficient, with in general nonlogarithmic contributions and logarithmic contributions.

Nevertheless without any ambiguity, numerically in the language of EFT, the most constraining $T$ parameter are taken into account in a general way, and comparison is made among all the most stringent operators. Assuming the SM like Higgs mass relation in the MSSM, the probed region for each future experiments are shown. In particular in the large $X_t$ branch the constraints can be pushed to very high values, and probably rule out the low scale stop sector, by precisions provided by, e.g., the FCC-ee experiment.

\acknowledgments

The author is grateful to Carlos E.M.~Wagner and Lian-Tao Wang for useful discussions, and especially to Minyuan Jiang for an independent check and comparison, pointing out omissions in the first version. This research was supported in part by the World Premier International Research Center Initiative, Ministry of Education, Culture, Sports, Science and Technology, Japan.

\appendix
\section{More on Feynman Parametrization and Its Integration\label{sec:app}}

After taking trace, the denominator of each term is a multiplication of $(p_E^2+M_i^2+u)$. Each term eventually contributing to dimension-6 operators in Table~\ref{tab:Op} actually falls into one of the following categories
\begin{align}
\frac{1}{\prod_{\sum n_i=4}(p_E^2+M_i^2+u)^{n_i}}&,&\frac{p_E^2}{\prod_{\sum n_i=5}(p_E^2+M_i^2+u)^{n_i}}&,&\nonumber\\
\frac{X_iX_j}{\prod_{\sum n_i=5}(p_E^2+M_i^2+u)^{n_i}}&,&\frac{p_E^2X_iX_j}{\prod_{\sum n_i=6}(p_E^2+M_i^2+u)^{n_i}}&,&\frac{p_E^4X_iX_j}{\prod_{\sum n_i=7}(p_E^2+M_i^2+u)^{n_i}},\nonumber\\
\frac{X_iX_jX_kX_l}{\prod_{\sum n_i=6}(p_E^2+M_i^2+u)^{n_i}}&,&\frac{p_E^2X_iX_jX_kX_l}{\prod_{\sum n_i=7}(p_E^2+M_i^2+u)^{n_i}}&,&\nonumber\\
\frac{X_iX_jX_kX_lX_mX_n}{\prod_{\sum n_i=7}(p_E^2+M_i^2+u)^{n_i}}&.&&&\label{eq:aftertrace}
\end{align}
Here $X_i$ generally indicates some dimension-1 quantity, which could be the $X_t,X_b$ terms in our sfermion sector, or the large scale itself in the vectorlike fermion model.

To see this is the case we should at first go through the following calculation steps. The standard textbook trick of Feynman parametrization allows the loop integration over $\int d^dp_E$, with dimensional regularization/reduction which makes no difference at one loop level. Then the integration over $\int du$ can always be performed, with the boundary $u\to\infty$ dropped as $\overline{\text{MS}}$ subtraction. The first integration increase the dimension by 4 and the second integration by 2, and all the above categories eventually give correct dimension of $-2$, as required for the dimension-6 operators. Note that each $p_E$ on the numerator corresponds to a $\textstyle\frac{\partial}{\partial p}$ and so that a covariant derivative in the $\delta\tilde{V}''+\tilde{G}$, checking with all possible combinatorics for each operator in Table~\ref{tab:Op} (see the Appendix of~\cite{Huo:2015exa}), the only $p_E^4$ term arise from the $\mathcal{O}_D$ calculation\footnote{The field strength related $p_E^4$ term such as in the combinatoric of $\frac{4}{3!}gp^\mu t^aD_\rho F_{\nu\mu}^a{\textstyle\frac{\partial}{\partial p}^\rho\frac{\partial}{\partial p}^\nu}$ always cancels due to the antisymmetric property of $F_{\nu\mu}^a$.} and no term falls into the category of $p_E^4/\prod_{\sum n_i=6}(p_E^2+M_i^2+u)^{n_i}$.

After finishing the loop integration and $\int du$, the Feynman parameter integration we actually need to do is
\begin{equation}
F((p^2)^{n_p},(M_1^2)^{n_1},(M_2^2)^{n_2},\cdots)=\frac{(\sum n_i-1)!}{\prod(n_i-1)!}\left(\prod\int_0^1dx_i\right)\delta(\sum x_i-1)\frac{\prod x_i^{n_i-1}}{(\sum M_i^2x_i)^{\sum n_i-n_p-3}}.
\end{equation}
Here we do not keep the factors from loop integration, and the $n_p$ and $n_1,n_2,\cdots$ corresponds to categories in Eq.~(\ref{eq:aftertrace}). In our model of sfermion there are at most three different large mass scales. For the case of two large mass squares we list them all
\begin{align}
\label{eq:twoFeyn1}
F((p^2)^0,(M_1^2)^3,(M_2^2)^1)&=\frac{3(M_1^2-3M_2^2)}{2(M_1^2-M_2^2)^2}+\frac{3M_2^4}{(M_1^2-M_2^2)^3}\ln\frac{M_1^2}{M_2^2},&\\
F((p^2)^0,(M_1^2)^2,(M_2^2)^2)&=\frac{3(M_1^2+M_2^2)}{(M_1^2-M_2^2)^2}-\frac{6M_1^2M_2^2}{(M_1^2-M_2^2)^3}\ln\frac{M_1^2}{M_2^2},&\\
F((p^2)^0,(M_1^2)^4,(M_2^2)^1)&=\frac{2(M_1^4-5M_1^2M_2^2-2M_2^4)}{M_1^2(M_1^2-M_2^2)^3}+\frac{12M_2^4}{(M_1^2-M_2^2)^4}\ln\frac{M_1^2}{M_2^2},&\\
F((p^2)^0,(M_1^2)^3,(M_2^2)^2)&=\frac{6(M_1^2+5M_2^2)}{(M_1^2-M_2^2)^3}-\frac{12M_2^2(2M_1^2+M_2^2)}{(M_1^2-M_2^2)^4}\ln\frac{M_1^2}{M_2^2},&\\
F((p^2)^0,(M_1^2)^5,(M_2^2)^1)&=\frac{5(M_1^6-7M_1^4M_2^2-7M_1^2M_2^4+M_2^6)}{2M_1^4(M_1^2-M_2^2)^4}+\frac{30M_2^4}{(M_1^2-M_2^2)^5}\ln\frac{M_1^2}{M_2^2},&\\
F((p^2)^0,(M_1^2)^4,(M_2^2)^2)&=\frac{10(M_1^2+10M_1^2M_2^2+M_2^4)}{M_1^2(M_1^2-M_2^2)^4}-\frac{60M_2^2(M_1^2+M_2^2)}{(M_1^2-M_2^2)^5}\ln\frac{M_1^2}{M_2^2},&\\
F((p^2)^0,(M_1^2)^3,(M_2^2)^3)&=-\frac{90(M_1^2+M_2^2)}{(M_1^2-M_2^2)^4}+\frac{30(M_1^4+4M_1^2M_2^2+M_2^4)}{(M_1^2-M_2^2)^5}\ln\frac{M_1^2}{M_2^2},&\\
F((p^2)^0,(M_1^2)^6,(M_2^2)^1)&=\frac{3M_1^8-27M_1^6M_2^2-47M_1^4M_2^4+13M_1^2M_2^6-2M_2^8}{M_1^6(M_1^2-M_2^2)^5}+\frac{60M_2^4}{(M_1^2-M_2^2)^6}\ln\frac{M_1^2}{M_2^2},&\\
F((p^2)^0,(M_1^2)^5,(M_2^2)^2)&=\frac{5(3M_1^6+47M_1^4M_2^2+11M_1^2M_2^4-M_2^6)}{M_1^4(M_1^2-M_2^2)^5}-\frac{60M_2^2(2M_1^2+3M_2^2)}{(M_1^2-M_2^2)^6}\ln\frac{M_1^2}{M_2^2},&\\
F((p^2)^0,(M_1^2)^4,(M_2^2)^3)&=-\frac{20(10M_1^4+19M_1^2M_2^2+M_2^4)}{M_1^2(M_1^2-M_2^2)^5}+\frac{60(M_1^4+6M_1^2M_2^2+3M_2^4)}{(M_1^2-M_2^2)^6}\ln\frac{M_1^2}{M_2^2},&\\
\nonumber\\
F((p^2)^1,(M_1^2)^4,(M_2^2)^1)&=\frac{2(2M_1^4-7M_1^2M_2^2+11M_2^4)}{3(M_1^2-M_2^2)^3}-\frac{4M_2^6}{(M_1^2-M_2^2)^4}\ln\frac{M_1^2}{M_2^2},&\\
F((p^2)^1,(M_1^2)^3,(M_2^2)^2)&=\frac{2(M_1^4-5M_1^2M_2^2-2M_2^4)}{(M_1^2-M_2^2)^3}+\frac{12M_1^2M_2^4}{(M_1^2-M_2^2)^4}\ln\frac{M_1^2}{M_2^2},&\\
F((p^2)^1,(M_1^2)^5,(M_2^2)^1)&=\frac{5(M_1^6-5M_1^4M_2^2+13M_1^2M_2^4+3M_2^6)}{3M_1^2(M_1^2-M_2^2)^4}-\frac{20M_2^6}{(M_1^2-M_2^2)^5}\ln\frac{M_1^2}{M_2^2},&\\
F((p^2)^1,(M_1^2)^4,(M_2^2)^2)&=\frac{10(M_1^4-8M_1^2M_2^2-17M_2^4)}{3(M_1^2-M_2^2)^4}+\frac{20M_2^4(3M_1^2+M_2^2)}{(M_1^2-M_2^2)^5}\ln\frac{M_1^2}{M_2^2},&\\
F((p^2)^1,(M_1^2)^3,(M_2^2)^3)&=\frac{10(M_1^4+10M_1^2M_2^2+M_2^4)}{(M_1^2-M_2^2)^4}-\frac{60M_1^2M_2^2(M_1^2+M_2^2)}{(M_1^2-M_2^2)^5}\ln\frac{M_1^2}{M_2^2},&\\
\nonumber\\
F((p^2)^2,(M_1^2)^5,(M_2^2)^1)&=\frac{5(3M_1^6-13M_1^4M_2^2+23M_1^2M_2^4-25M_2^6)}{12(M_1^2-M_2^2)^4}+\frac{5M_2^8}{(M_1^2-M_2^2)^5}\ln\frac{M_1^2}{M_2^2},&\\
F((p^2)^2,(M_1^2)^4,(M_2^2)^2)&=\frac{5(M_1^6-5M_1^4M_2^2+13M_1^2M_2^4+3M_2^6)}{3(M_1^2-M_2^2)^4}-\frac{20M_1^2M_2^6}{(M_1^2-M_2^2)^5}\ln\frac{M_1^2}{M_2^2},&\\
F((p^2)^2,(M_1^2)^3,(M_2^2)^3)&=\frac{5(M_1^6-7M_1^4M_2^2-7M_1^2M_2^4+M_2^6)}{2(M_1^2-M_2^2)^4}-\frac{30M_1^4M_2^4}{(M_1^2-M_2^2)^5}\ln\frac{M_1^2}{M_2^2},&\\
F((p^2)^2,(M_1^2)^6,(M_2^2)^1)&=\frac{3M_1^8-17M_1^6M_2^2+43M_1^4M_2^4-77M_1^2M_2^6-12M_2^8}{2M_1^2(M_1^2-M_2^2)^5}+\frac{30M_2^8}{(M_1^2-M_2^2)^6}\ln\frac{M_1^2}{M_2^2},&\\
F((p^2)^2,(M_1^2)^5,(M_2^2)^2)&=\frac{5(M_1^6-7M_1^4M_2^2+29M_1^2M_2^4+37M_2^6)}{2(M_1^2-M_2^2)^5}-\frac{30M_2^6(4M_1^2+M_2^2)}{(M_1^2-M_2^2)^6}\ln\frac{M_1^2}{M_2^2},&\\
F((p^2)^2,(M_1^2)^4,(M_2^2)^3)&=\frac{5(M_1^6-11M_1^4M_2^2-47M_1^2M_2^4-3M_2^6)}{(M_1^2-M_2^2)^5}+\frac{60M_1^2M_2^4(3M_1^2+2M_2^2)}{(M_1^2-M_2^2)^6}\ln\frac{M_1^2}{M_2^2}.&
\label{eq:twoFeyn2}
\end{align}
The three different large mass scales case can be worked out similarly.


\end{document}